\documentclass[twocolumn]{aastex631}

\usepackage[dvipsnames]{xcolor}
\definecolor{MyBlue}{RGB}{0,0,255}
\usepackage{hyperref}
\usepackage{subfigure}
\usepackage{longtable}
\usepackage{makecell}
\usepackage{tabularx}
\usepackage{amsmath}
\usepackage{enumitem}
\setlist{nosep}

\begin{document}

\title{X-ray Flare Evolution and Changing-Look Nature of the TeV Blazar OQ 334}

\author[0000-0001-9092-2903]{P. K. Navaneeth}
\affiliation{Doctoral School of Exact and Technical Sciences, University of Zielona Góra, 65-516 Zielona Góra, Poland; email: \textcolor{blue}{940d@g.elearn.uz.zgora.pl}}

\author[0000-0002-0705-6619]{Gopal Bhatta}
\affiliation{Janusz Gil Institute of Astronomy, University of Zielona Góra, 65-516 Zielona Góra, Poland; email: \textcolor{blue}{g.bhatta@ia.uz.zgora.pl}}

\author[0000-0001-8375-1907]{Cosimo Nigro}
\affiliation{Institut de Física dAltes Energies (IFAE), The Barcelona Institute of Science and Technology, Campus UAB, 08193 Bellaterra (Barcelona), Spain; email: \textcolor{blue}{cosimo.nigro@ifae.es}}
\affiliation{Departament de Física, Universitat Autònoma de Barcelona, 08193 Bellaterra, Spain}

\begin{abstract}
The blazar OQ 334 (B2 1420+326, z = 0.682) is one of the few jetted AGN with repeated changing-look (CL) transitions between FSRQ and BL Lac states. We present a multi-wavelength study covering 2017 December to 2024 September, focusing on three X-ray flares observed with Swift-XRT (X1: 2019 June 26--July 24, X2: 2020 January 4--February 21, X3: 2024 June 24--August 2) and on the spectral properties of the CL states. The spectrum hardens from flare to flare, with $\Gamma$ decreasing from $1.97 \pm 0.09$ to $1.71 \pm 0.04$ and $1.34 \pm 0.06$, while the hysteresis evolves from an initial clockwise loop, consistent with cooling-dominated emission, to multi-loop trajectories in the later flares. The X-ray spectra separate the CL states, with FSRQ states hardest ($\Gamma_X \approx 1.3$--$1.5$), BL Lac states softest ($\Gamma_X \approx 1.8$--$2.4$), and transition states intermediate. Within the CL states, flux and photon index are positively correlated ($r = 0.59$, $p = 2.5 \times 10^{-2}$), producing a softer-when-brighter trend. The first two flares coincide with FSRQ $\rightarrow$ BL Lac transitions, whereas the third, the brightest, remains FSRQ-like. One-zone leptonic modelling of the broadband SEDs reproduces the flares with a weak, nearly constant magnetic field and a particle-dominated region, and the X-ray emission evolves from a substantial synchrotron contribution in the first flare to complete inverse-Compton dominance in the third. The X-ray band traces the changing balance between the synchrotron and inverse-Compton components in the flares and provides a sensitive probe of the CL behaviour of the source.
\end{abstract}

\keywords{Blazars, Flat-spectrum radio quasars, BL Lacertae objects, Relativistic jets, High energy astrophysics}

\section{Introduction} \label{sec:introduction}
Blazars are a subclass of radio-loud active galactic nuclei (AGN) powered by a central supermassive black hole, characterised by highly relativistic jets directed at small angles to the observer's line of sight \citep{Urry1995, Blandford1977}. Their emission is non-thermal, spanning the entire electromagnetic spectrum from radio to very-high-energy (VHE, $E \gtrsim 100\,\mathrm{GeV}$) $\gamma$-rays, and is significantly enhanced by relativistic beaming \citep{Blandford1979}. Blazars are among the most luminous persistent sources in the universe, dominating the extragalactic $\gamma$-ray sky. They exhibit extreme observational properties, including strong flux variability across all wavelengths on timescales ranging from minutes to years \citep{Heidt1996, Bhatta2021}, high optical polarisation, superluminal motion, and strong $\gamma$-ray emission \citep{Miller1978, Hartman1999, Abdo2010}. 

The broadband spectral energy distribution (SED) of blazars displays a characteristic double-humped structure. The low-energy hump, spanning from radio to optical/X-ray frequencies, is  attributed to synchrotron radiation from relativistic electrons in the jet \citep{Urry1998}. The origin of the high-energy hump, peaking in the MeV-TeV range, remains a subject of active investigation, with two broad classes of models proposed. In the leptonic framework, this emission is produced via inverse-Compton (IC) scattering of low-energy photons by the relativistic electron population in the jet. The seed photons for this process may be internal to the jet, as in the synchrotron self-Compton (SSC) model \citep{Maraschi1992, Marscher1985}, or may originate from external photon fields such as the accretion disk, the broad-line region (BLR), or the dusty torus, as in the external Compton (EC) model \citep{Dermer1992, Sikora1994, Blandford1995, Ghisellini1996}. Hadronic models, on the other hand, attribute the high-energy emission to proton synchrotron radiation or to secondary particles produced in photo-pion and photo-pair interactions \citep{Aharonian2002, Bottcher2013}. 

Blazars are generally classified in two ways: by the position of their synchrotron peak and by their optical/UV emissions. Based on the rest-frame equivalent width (EW) of their broad emission lines (BELs), they are divided into Flat Spectrum Radio Quasars (FSRQs), which exhibit strong lines with $\mathrm{EW} > 5\,\text{\AA}$, and BL Lacertae objects (BL Lacs), which show featureless spectra or weak lines with $\mathrm{EW} < 5\,\text{\AA}$ \citep{Stocke1991,Urry1995}. This observational distinction often corresponds to different accretion regimes, where FSRQs are associated with radiatively efficient accretion and BL Lacs with radiatively inefficient modes \citep{Ghisellini2001, Ghisellini2011}. Blazars are also categorised by the frequency of their low-energy synchrotron peak ($\nu_{\mathrm{p}}$). They are classified as Low Synchrotron Peaked (LSPs) when $\nu_{\mathrm{p}} \leq 10^{14}\,\mathrm{Hz}$, Intermediate Synchrotron Peaked (ISPs) when $10^{14}\,\mathrm{Hz} < \nu_{\mathrm{p}} < 10^{15}\,\mathrm{Hz}$, and High Synchrotron Peaked (HSPs) when $\nu_{\mathrm{p}} \geq 10^{15}\,\mathrm{Hz}$ \citep{Abdo2010}.

The discovery of changing-look blazars (CLBs) has challenged these traditional categories, as these transitional sources switch between FSRQ and BL Lac states over timescales of months to years \citep{Vermeulen1995, Ghisellini2011b, Ghisellini2013, Ruan2014, Mishra2021, PenaHerazo2021, Ricci2023, Paiano2024}. These transitions, such as those observed in OQ 334, involve dramatic variations in the $\mathrm{EW}$ of BELs (\citealp{Mishra2021}). The physical mechanisms triggering these changes are still under investigation, with primary theories suggesting either sudden shifts in the accretion rate \citep{Feng2021, Xiao2022, Duffy2025} or the dilution of emission lines by a highly variable, Doppler-boosted jet continuum during active flares \citep{Blandford1978, Ruan2014}. Other recent studies propose that CLBs represent a specific, short-lived phase in the evolutionary process of blazars \citep{Lyu2022, Kang2023, Ren2024a}.

OQ 334, also referred to as B2 1420+32 or QSO B1420+326 (RA: 215.6265, Dec: 32.3862), is a blazar located at a redshift of $z = 0.682$ \citep{Hewett2010}. Based on its radio spectrum, the source has been historically classified as an FSRQ \citep{Healey2007} and is listed as such in the Fermi-LAT Fourth Source Catalogue \citep[4FGL;][]{Abdollahi2020}. The central black hole has an estimated mass of $M_\mathrm{BH} \simeq 4 \times 10^8\,M_\odot$, inferred from its luminosity and broad-line width \citep{Brotherton2015}. After nearly a decade of relative quiescence, the source underwent a dramatic optical flare of more than 2 magnitudes in December 2017 \citep{Mishra2021}, marking the beginning of a prolonged period of elevated activity across the electromagnetic spectrum, including optical, infrared, and $\gamma$-ray bands \citep{Ciprini2018, Mishra2021, Prince2021}. During this active period, \citet{Mishra2021} found evidence of repeated transitions between the FSRQ and BL Lac states, beginning in January 2018, establishing OQ 334 as a changing-look blazar. The continued high state in late 2019 triggered follow-up observations by the MAGIC telescopes, leading to the discovery of VHE $\gamma$-ray emission in January 2020 \citep{Mirzoyan2020, MAGICCollaboration2021}.

\begin{figure*}[!t]
    \centering
    \includegraphics[width=\textwidth,height=0.75\textheight,keepaspectratio]{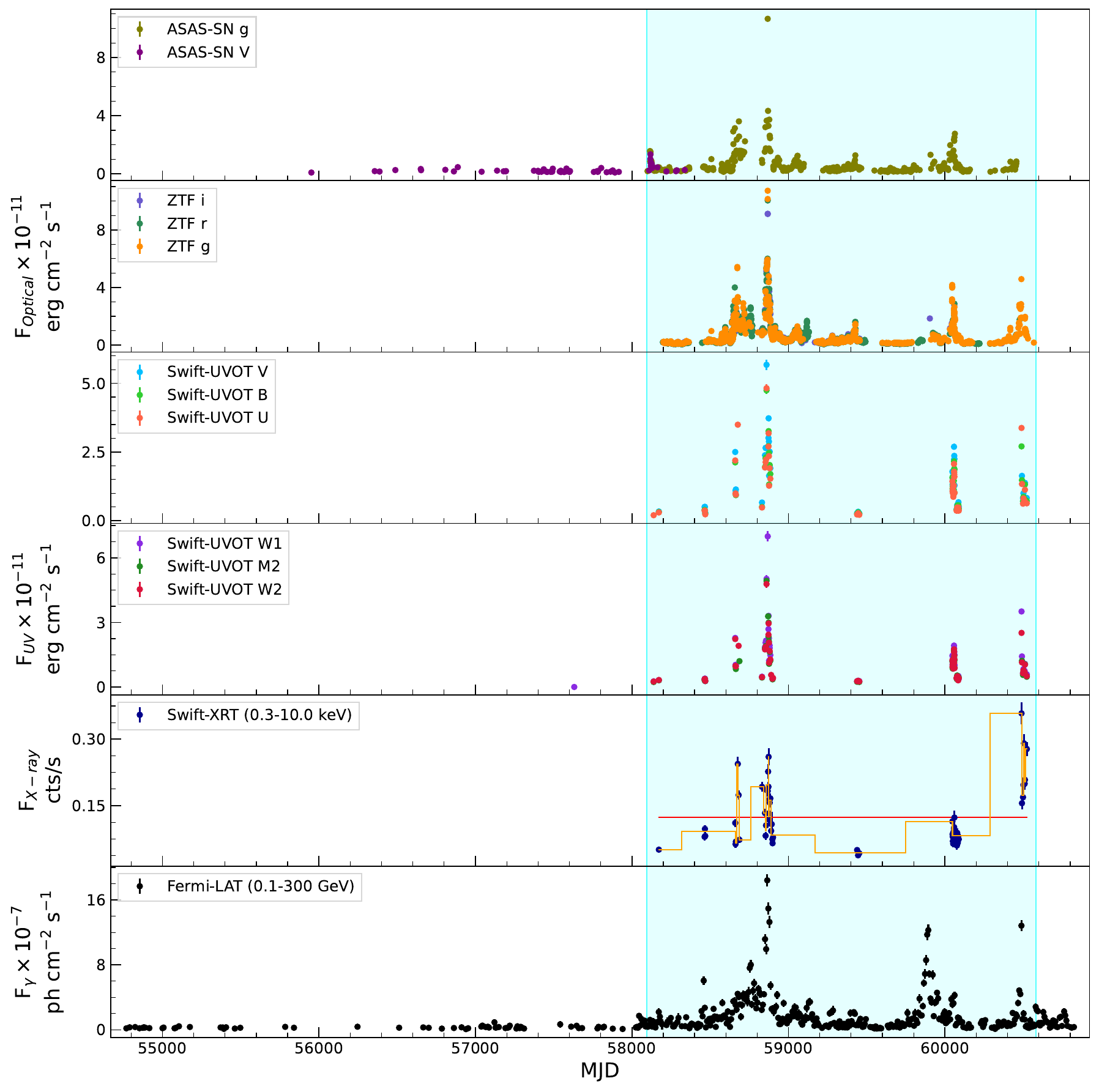}
    \caption{Multi-wavelength light curves of the FSRQ OQ 334 (B2 1420+326) from MJD 54769.0 to 60824.0. From top to bottom, the panels show: ASAS-SN optical, ZTF optical, Swift-UVOT optical, Swift-UVOT Ultraviolet, Swift-XRT X-ray, and Fermi-LAT $\gamma$-ray observations. The shaded region indicates the time period analysed in this work (MJD 58095.0–60580.0). In the Swift-XRT panel, the orange lines mark the change points identified using Bayesian-block analysis, while the red horizontal line indicates the mean X-ray flux over the analysed period.}
    \label{fig:mwllc}
\end{figure*}

The paper is organised as follows. In Section~\ref{sec:data}, we describe the multi-wavelength dataset and associated data reduction procedures. Section~\ref{sec:analysis} presents the analysis and results, including the timing and optical analysis, X-ray spectral analysis, changing-look nature, and broadband SED modelling. In Section~\ref{sec:discussion}, we discuss the physical implications of our findings, and we summarise our conclusions in Section~\ref{sec:conclusions}.

\section{Data Reduction} \label{sec:data}

\subsection{MAGIC}
MAGIC is a stereoscopic system consisting of two 17\,m diameter ground-based Imaging Atmospheric Cherenkov Telescopes (IACTs) located on the island of La Palma, Spain \citep{Aleksic2016}. The telescopes record Cherenkov light from extensive air showers initiated by gamma rays, covering an energy range from ${\sim}50$ GeV to ${\sim}100$ TeV. With an integral sensitivity of $(0.66 \pm 0.03)\%$ of the Crab Nebula flux above 220\,GeV for point-like sources with a Crab Nebula-like spectrum, achieved in 50 h of observations \citep{Aleksic2016a}, MAGIC is among the most sensitive instruments for VHE blazar studies.

\begin{figure}
    \centering
    \includegraphics[width=0.9\linewidth]{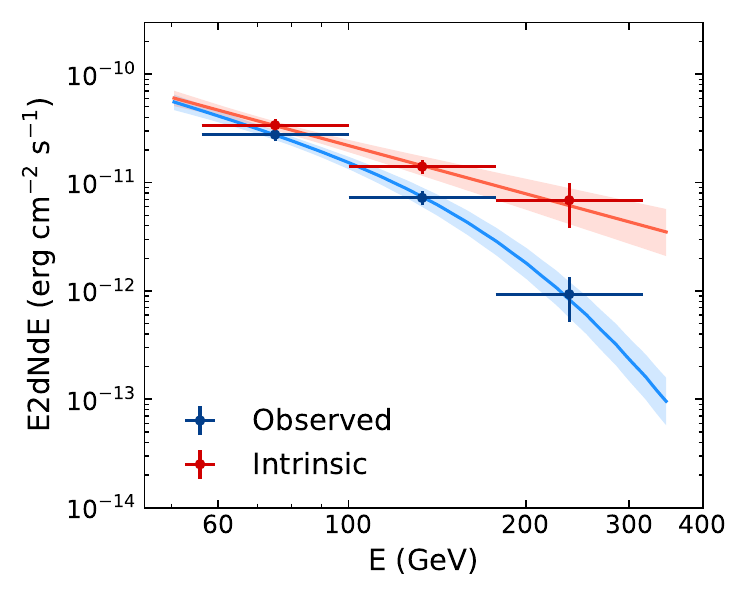}
    \caption{MAGIC very high-energy $\gamma$-ray spectra of OQ 334 during the flare period MJD 58869.22-58885.25. The observed and intrinsic spectra are shown together with their respective fits. The intrinsic spectrum has been corrected for attenuation by the extragalactic background light, and the shaded bands indicate the statistical uncertainties of the corresponding fits.}
    \label{fig:magic_sed}
\end{figure}

The observations analysed span MJD $58869.22-58885.25$, corresponding to periods C (VHE flare) and D (post-flare) reported by \citet{MAGICCollaboration2021}. Calibration, image cleaning, and event reconstruction were performed with the MARS package \citep{Zanin2013oib}, after which the analysis was carried out from the DL3 level \citep{Abe2024Nov} using \texttt{Gammapy} \citep{Donath2023}. The signal was extracted from a point-like region centred on the source, with the extraction radius taken from the energy-dependent \texttt{RAD\_MAX} values stored in the IRFs, and the background was estimated using three reflected regions placed at the same camera offset. The individual observations were stacked into a single ON-OFF dataset, and the fit was restricted to the $50-500$ GeV range. The intrinsic spectrum was modelled with a power law,
\begin{equation}
    \frac{\mathrm{d}N}{\mathrm{d}E} = N_{0} \left( \frac{E}{E_{0}} \right)^{-\Gamma} ,
    \label{eq:magic_pwl}
\end{equation}
multiplied by an extragalactic background light (EBL) absorption term $e^{-\tau(E,z)}$ following the model of \citet{Domnguez1102}. The normalisation and photon index were left free, while the reference energy ($E_{0} = 0.1$ TeV) and redshift were fixed. The fit gives $N_{0} = (1.37 \pm 0.18) \times 10^{-9}$\,TeV$^{-1}$\,cm$^{-2}$\,s$^{-1}$ and $\Gamma = 3.48 \pm 0.32$. The intrinsic points were obtained by referencing to the unabsorbed power law. The intrinsic and observed spectra of the source are shown in Figure~\ref{fig:magic_sed}.

\subsection{Fermi}
The Fermi $\gamma$-ray Space Telescope, launched by NASA on June 11, 2008, is designed to study the high-energy gamma-ray sky. The observatory carries two instruments onboard: the Large Area Telescope (LAT) and the Gamma-ray Burst Monitor (GBM), together enabling gamma-ray observations spanning from a 8 keV to beyond 300 GeV \citep{Atwood2009, Meegan2009}. The LAT is the primary instrument, a pair-conversion telescope designed to detect gamma rays in the energy range of $\sim$20 MeV to more than 300 GeV, featuring a large effective area of $>$8000 cm$^2$ at $\sim$1 GeV and a wide field of view of 2.4 sr \citep{Atwood2009}. Operating in survey mode, the LAT covers approximately 20\% of the sky at any given time and scans the entire sky every $\sim$3 hours, making it exceptionally well-suited for monitoring the long-term gamma-ray variability and spectral properties of blazars over a broad range of timescales.

The LAT data were downloaded from the Fermi Science Support Center\footnote{\url{https://fermi.gsfc.nasa.gov/ssc/data/access/lat/}} for a 20$^{\circ}$ region of interest (ROI) centred on the source position. The analysis utilised Pass 8 (P8R3) data \citep{Atwood2013} to search for gamma-ray emission in the energy range of 0.1-300 GeV, employing FermiTools 2.2.0 and Fermipy 1.2.2 \citep{Wood2017}. The SOURCE class events (evclass = 128 and evtype = 3) were selected within a $10^{\circ}$ ROI centred on the source position, adopting the P8R3\_SOURCE\_V3 instrument response functions. Standard data quality cuts were applied, including a zenith angle cut of $z_{\rm max} < 90^{\circ}$ to reduce contamination from the Earth's limb, and the selection of good time intervals based on \texttt{(DATA\_QUAL>0)\&\&(LAT\_CONFIG==1)}. The significance of the $\gamma$-ray signal was evaluated using a maximum-likelihood test statistic (TS) defined as TS $= 2(\log\mathcal{L}_1 -\log\mathcal{L}_0)$, where $\mathcal{L}_1$ and $\mathcal{L}_0$ are the likelihoods with and without a point source at the position of the target \citep{Mattox1996}. All point sources from the 4FGL catalog \citep{Abdollahi2020} within $15^{\circ}$ of the ROI centre were included in the initial model file, along with the Galactic (\texttt{gll\_iem\_v07}) and extragalactic isotropic diffuse emission (\texttt{iso\_P8R3\_SOURCE\_V3\_v1}) components. 

A binned likelihood analysis was performed using eight energy bins per decade and a spatial binning of $0.1^{\circ}$ per pixel to derive the spectral parameters. After initial optimisation, sources with TS $<$ 10 were fixed, and the normalisation of sources with TS $>$ 10 was allowed to vary in the model file. The normalisation parameters for all sources within $3^{\circ}$ of the ROI centre were additionally freed, and all parameters associated with the isotropic and Galactic diffuse components were allowed to vary freely. To investigate the temporal variability of the source, a weekly binned light curve was generated. In each time bin, the normalisation parameters of all sources within $3^{\circ}$ of the ROI centre were left free to vary, along with the diffuse components, while the spectral parameters of all remaining sources were fixed to the values obtained from the global fit. Only time bins with TS $>$ 9 were considered statistically significant detections and retained in the light curve.

\subsection{Swift}
Swift is a multi-wavelength observatory launched in 2004 by NASA for Gamma-Ray Burst (GRB) astronomy, which is a medium-sized explorer (MIDEX) mission including three telescopes: Burst Alert Telescope (BAT), X-Ray Telescope (XRT), and Ultraviolet/Optical Telescope (UVOT) \citep{Gehrels2004}. While primarily designed for studying GRBs, Swift has proven valuable for multi-wavelength studies of various compact objects and facilitates the detection and analysis of diverse sources in high-energy astrophysics. The observatory's capability for rapid slewing and simultaneous multi-wavelength observations makes it invaluable for studying the highly variable nature of blazars across ultraviolet, optical, and X-ray bands.

The XRT is a focusing X-ray imaging spectrometer with $>120\ \rm{cm}^2$ effective area at 1.5 keV, $23^\prime$ field of view, and $18^{\prime\prime}$ angular resolution (half-power diameter), operating across the 0.2-10 keV energy range \citep{Burrows2005}. The telescope provides sensitive X-ray spectroscopy and photometry capabilities through an auto-exposure mode that dynamically adjusts readout mode to optimise data quality as source flux varies. The X-ray spectra and light curve of the source were generated using the Swift-XRT Online Data Products Generator tool\footnote{\url{https://www.swift.ac.uk/user_objects/}}. The data reduction process followed by this online tool can be found in \cite{Evans2007, Evans2009}.  Due to the low number of counts, the 0.3-10 keV X-ray spectra were grouped to a minimum of one count per bin using the \texttt{GRPPHA} task of FTOOLS. The spectra were then fitted in XSPEC version 12.14.0 \citep{Arnaud1996} using Cash statistics \citep{Cash1979}.

The UVOT has a modified Ritchey-Chrétien telescope design with a 30 cm diameter aperture, featuring 11 filters that cover a wavelength range of 170-600 nm across a 17$\times$17 arcmin$^2$ field of view \citep{Roming2005}. The source is observed with six standard photometric filters (Optical: B, V, U; UV: UVM2, UVW1, UVW2). The observations were downloaded from the SWIFTMASTER catalog and data analysis was performed using the HEASoft package version 6.33 and the CALDB version 20220331. The data reduction was performed by following the standard procedures in the Swift UVOT Software Guide\footnote{\url{https://www.swift.ac.uk/analysis/uvot}}. The source counts were extracted using a $6^{\prime\prime}$ radius centred on the source, while background counts were measured from a $20^{\prime\prime}$ radius circle placed in a nearby source-free region. Images across all extensions were combined for each observation using \texttt{uvotimsum}, with processing performed independently for each filter. Aperture photometry was subsequently conducted using \texttt{uvotsource} for individual filters within each observation. Galactic extinction corrections were applied to the magnitudes using E(B-V) value $(1.07 \pm 0.08)\times 10^{-2}$ \citep{Schlafly2011, IRSA2022Jan} and the extinction laws of \citet{Cardelli1989}, and the corrected magnitudes were transformed into flux densities using zero points from \citet{Breeveld2011} and conversion factors from \citet{Poole2007}. 

\subsection{Optical Archival Data}
The All-Sky Automated Survey for Supernovae (ASAS-SN) \citep{Shappee2014, Hart2023} is a global network of 24 telescopes that has been routinely monitoring the entire visible sky since 2012. The survey operated in the V-band until late 2018, after which it transitioned to g-band observations. The archival optical light curve data for the source were retrieved from the ASAS-SN Sky Patrol\footnote{http://asas-sn.ifa.hawaii.edu/skypatrol} V2.0 Photometry Database.

The Zwicky Transient Facility (ZTF) \citep{Masci2018, Bellm2018} is a high-cadence optical time-domain survey operating on the Palomar 48-inch Schmidt telescope since 2018. ZTF observes the northern sky in three optical bands (g, r, and i), reaching typical depths of $\sim$20.5 mag with a nominal cadence of 3 days for fields in the main survey area. Archival optical light curve data for the source were obtained from the ZTF Data Release\footnote{https://www.ztf.caltech.edu/} using point spread function (PSF) fitting photometry. 

\begin{figure*}
\centering
\subfigure{\includegraphics[width=0.3\textwidth]{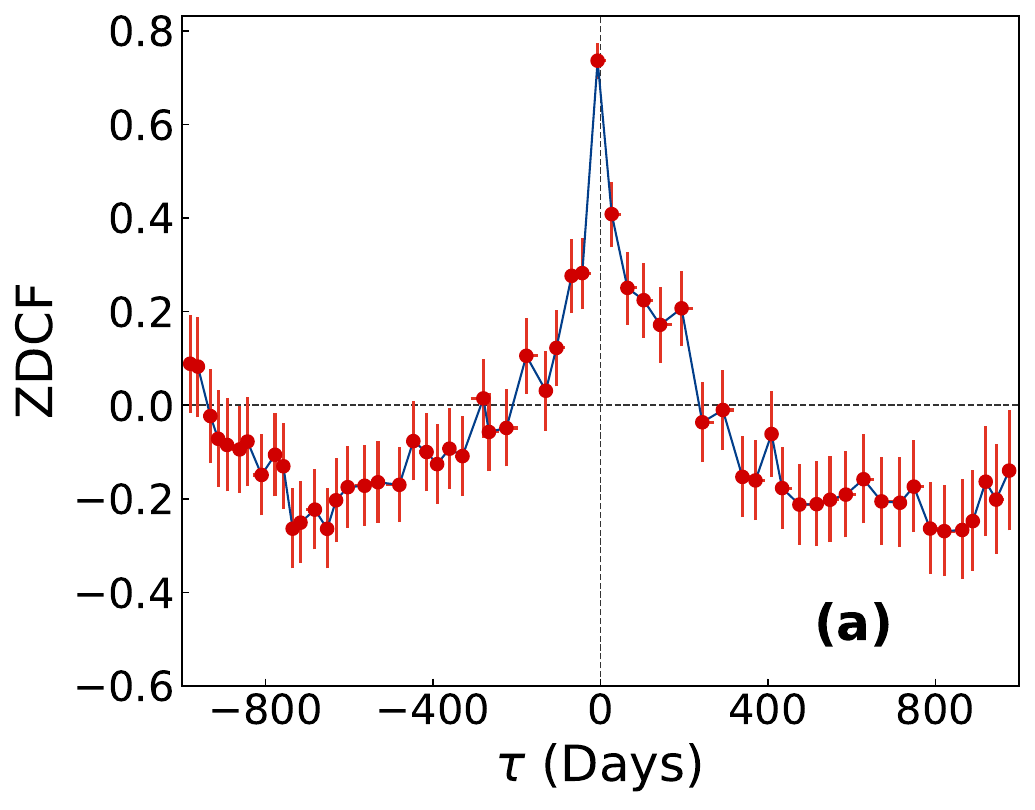}}
\subfigure{\includegraphics[width=0.3\textwidth]{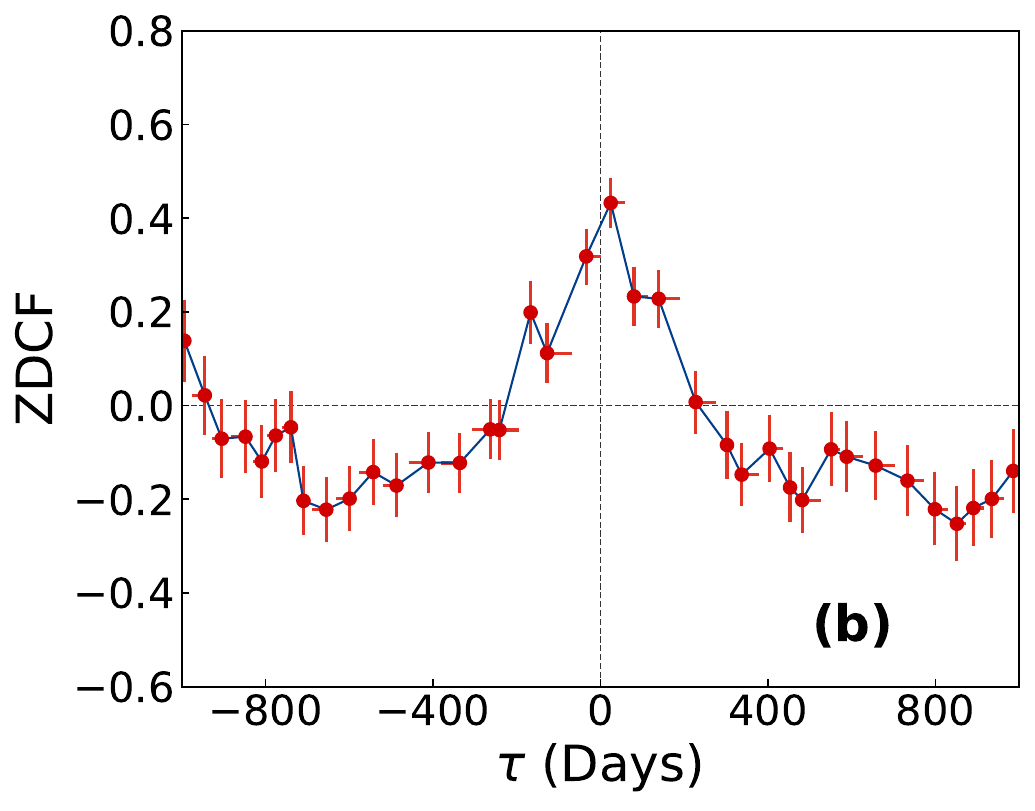}}
\subfigure{\includegraphics[width=0.3\textwidth]{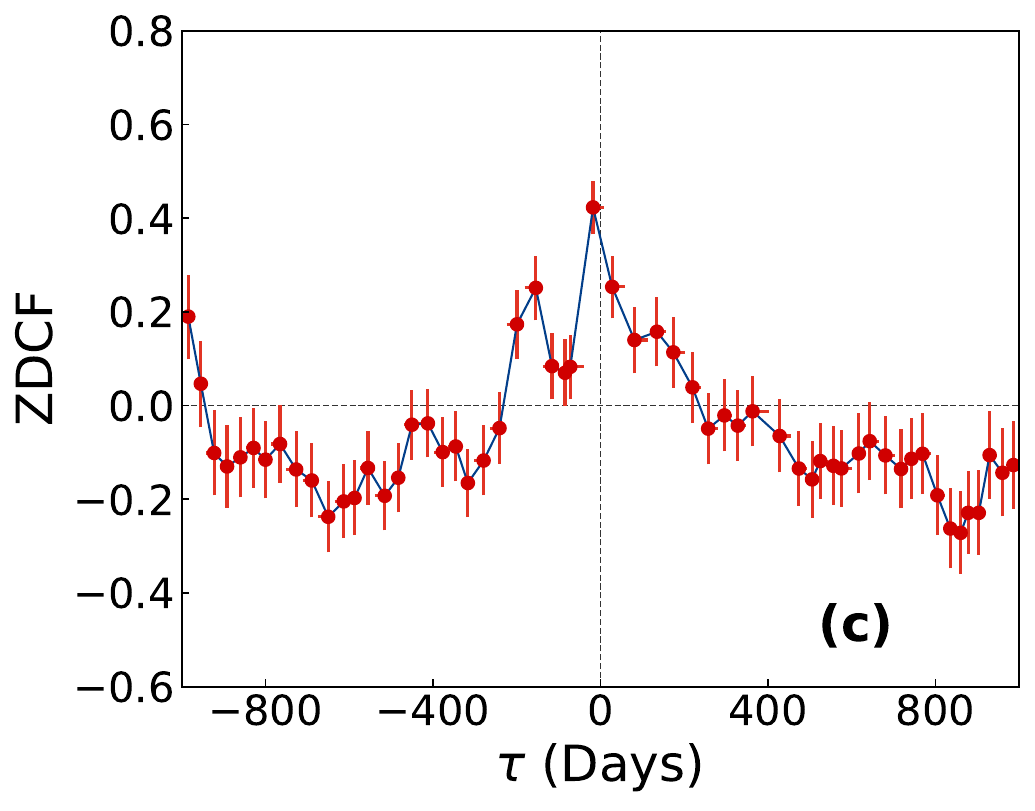}}
\caption{$z$-transformed discrete correlation functions (ZDCF) for OQ~334 between pairs of light curves across multiple wavebands. Panels (a)-(c) show the cross-correlation of the Fermi-LAT light curve with the ZTF $i$-band, ZTF $r$-band, and ASAS-SN $g$-band optical light curves, respectively.}
\label{fig:zdcf}
\end{figure*}

While ASAS-SN and ZTF were primarily designed to discover bright optical transients and identify variable phenomena, their continuous all-sky coverage, systematic observing cadence, and photometric precision make them well-suited for optical variability studies of blazars. The downloaded light curve data were filtered using instrument-specific quality cuts, corrected the magnitudes for Galactic extinction adopting the E(B-V) value $(1.07 \pm 0.08)\times 10^{-2}$ \citep{Schlafly2011, IRSA2022Jan} and extinction laws from \citet{Cardelli1989}, and converted the corrected magnitudes into flux densities.

\section{Analysis \& Results} \label{sec:analysis}
The multi-wavelength light curves of OQ 334 spanning MJD 54769.0 to 60824.0 is presented in Figure \ref{fig:mwllc}. For this work, we use the period from MJD 58095.0 to 60580.0, during which multi-wavelength data are available across all bands. Prior to MJD 58095.0, the source was in a quiescent state, as shown by Fermi-LAT and ASAS-SN V-band observations.

\subsection{Timing \& Optical Analysis} \label{sec:timing}
To quantify the inter-band correlations and the associated time lags, we performed a cross-correlation analysis of the optical and gamma-ray light curves. The Swift XRT and UVOT light curves were not included, as their sampling is too sparse for a meaningful correlation analysis.
The $z$-transformed Discrete Correlation Function (ZDCF; \citealt{Alexander9701, Alexander1302}) was used, which is a modification of the standard Discrete Correlation Function (DCF; \citealt{Edelson8810}; see also \citealt{Bhatta2112}). The ZDCF applies Fisher's $z$-transformation to the correlation coefficient $r$ to stabilise its skewed sampling distribution \citep{Alexander9701}. The correlation is computed from the unbinned discrete correlation function (UDCF) for each flux measurement pair $(x_i, y_j)$:
\begin{equation}
    \mathrm{UDCF}_{ij} = \frac{(x_i - \bar{x})(y_j - \bar{y})}{\sqrt{(S_x^{2} - \sigma^{2}_{\rm err,\mathit{x}})(S_y^{2} - \sigma^{2}_{\rm err,\mathit{y}})}},
\end{equation}
where $\bar{x}$ and $\bar{y}$ are the mean fluxes, $S_x^{2}$ and $S_y^{2}$ are their respective variances, and $\sigma_{\rm err,\mathit{x}}$ and $\sigma_{\rm err,\mathit{y}}$ are the flux uncertainties. These UDCF pairs are subsequently binned in equal-width time bins to obtain the DCF:
\begin{equation}
    \mathrm{DCF}(\tau) = \frac{1}{M} \sum \mathrm{UDCF}_{ij},
\end{equation}
where $M$ is the number of pairs within the bin.

The ZDCF analysis was implemented using the \texttt{pyZDCF} Python package \citep{Jankov2210}, and results are presented in Figure \ref{fig:zdcf}. The cross-correlation between the Fermi-LAT light curve and three optical bands (ZTF $i$, ZTF $r$, and ASAS-SN $g$) reveals a consistent peak near zero time lag. The peak occurs at $\tau \approx -7$ days for the LAT-ZTF $i$ pair, with DCF $= 0.74 \pm 0.04$, and at comparable lags for ZTF $r$ and ASAS-SN $g$, though those pairs do not exceed the DCF $> 0.5$ threshold.

\begin{figure*}
\centering
\subfigure{\includegraphics[width=0.3\textwidth]{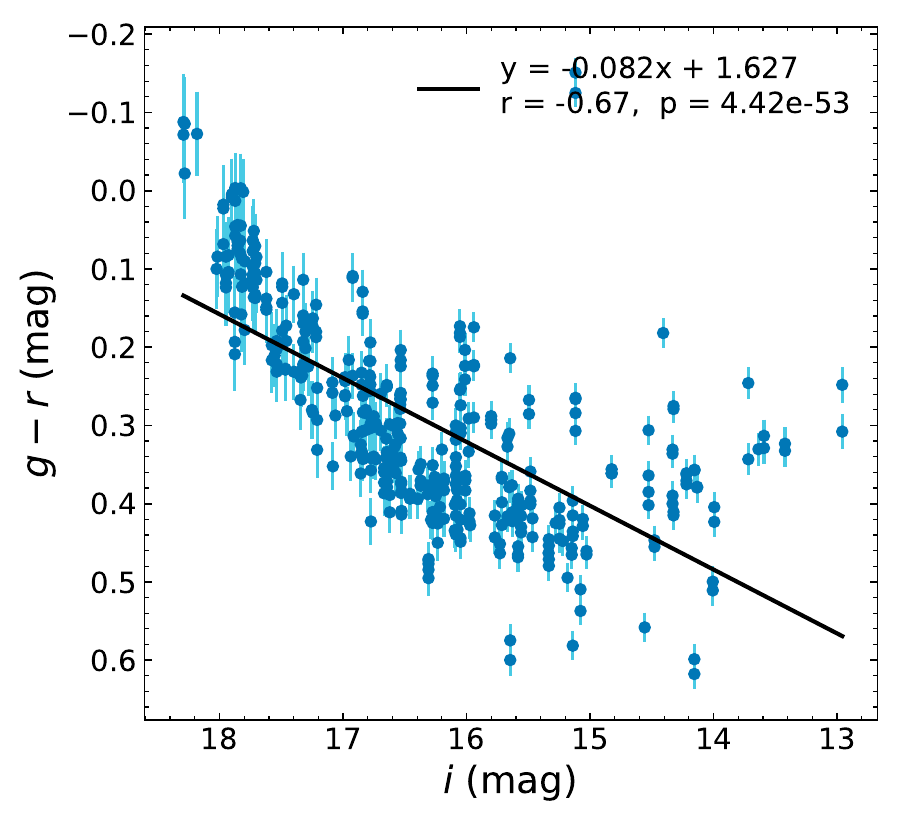}}
\subfigure{\includegraphics[width=0.3\textwidth]{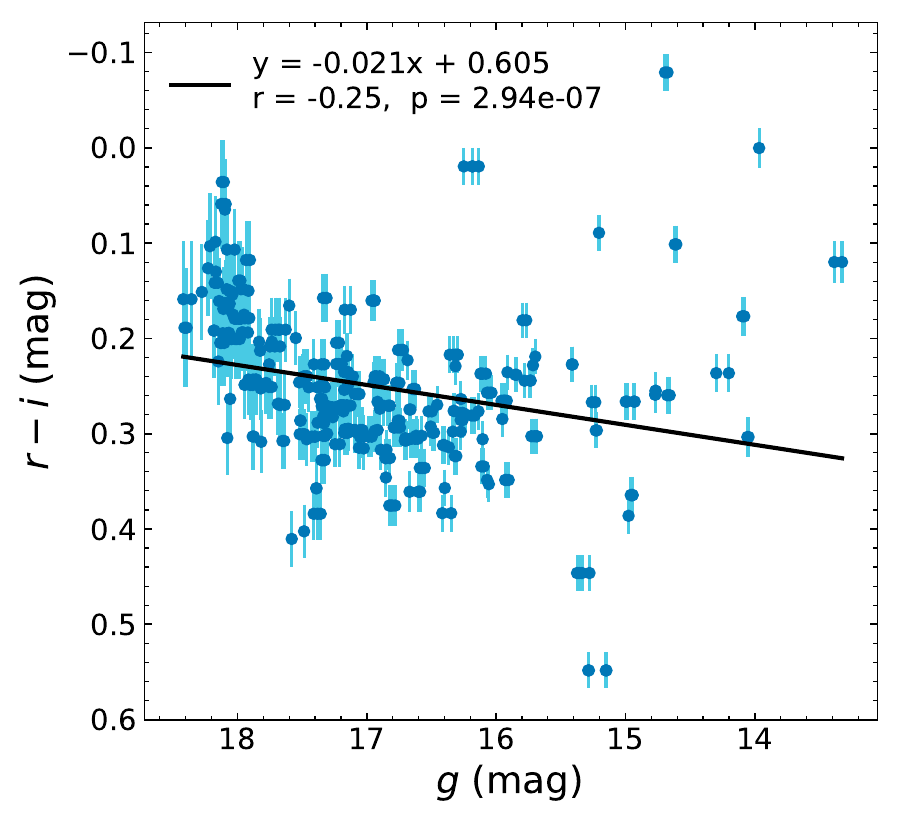}}
\subfigure{\includegraphics[width=0.3\textwidth]{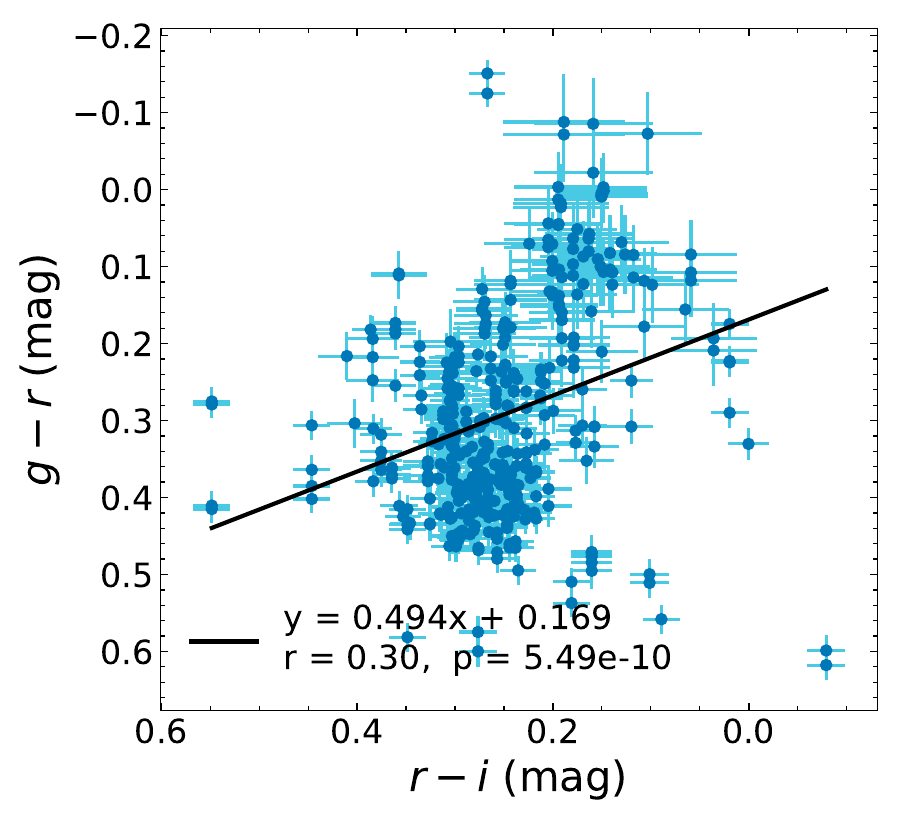}}
\caption{Colour-magnitude and colour-colour diagrams for OQ 334 constructed from ZTF $g$, $r$, and $i$ band photometry. Left and middle panels show the $g-r$ colour against the $i$ band magnitude and the $r-i$ colour against the $g$ band magnitude, respectively. The right panel shows $g-r$ against $r-i$. Solid lines are linear least-squares fits, Pearson correlation coefficient $r$, and the corresponding $p$ value given in each panel.}
\label{fig:ztf}
\end{figure*}

The spectral behaviour of the optical emission is studied using colour-magnitude and colour-colour diagrams in ZTF $g$, $r$, and $i$ bands, shown in Figure \ref{fig:ztf}. Each colour is plotted against the magnitude of the band not entering it, so that the uncertainties on the two axes remain independent. The $g-r$ colour is correlated with the $i$ band magnitude, with a Pearson coefficient of $-0.67$ and $p = 4.4 \times 10^{-53}$. The $r-i$ colour shows the same trend against the $g$ band magnitude, though more weakly, with a coefficient of $-0.25$ and $p = 2.9 \times 10^{-7}$. In both cases, the colour index decreases as the magnitude increases, and the optical spectrum becomes redder as the source brightens \citep{Chand2025}. The colour-colour diagram shows a positive correlation between $g-r$ and $r-i$, with a correlation coefficient of $0.30$, and $p = 5.5 \times 10^{-10}$.

\subsection{X-ray Analysis}

\subsubsection{Spectral Analysis}
A total of 64 Swift-XRT observations were available during the studied time period. We performed spectral fitting in the 0.3-10 keV energy range using three different absorbed (\texttt{tbabs}) spectral models: power-law, log-parabola, and broken power-law. When more than one model is used to fit a source, a statistical comparison among those models is required to select the best result. To identify the best fit for each observation, model comparison was carried out using the corrected Akaike Information Criterion \citep[$\text{AIC}_c$;][]{Akaike1974, Sugiura1978, Liddle2007}, defined as
\begin{equation}
    \text{AIC}_c = \text{AIC} + \frac{2k(k+1)}{N-k-1}
\end{equation}
where $\text{AIC} = 2k + C$, $C = -2\:\ln(L)$ is the Cash statistic value, $L$ is the maximum likelihood for the model, $k$ is the number of free parameters, and $N$ is the number of spectral bins. Given the relatively small number of spectral bins in some observations, the finite-sample correction is necessary \citep{Liddle2007}.

\begin{figure}
    \centering
    \includegraphics[width=0.9\linewidth]{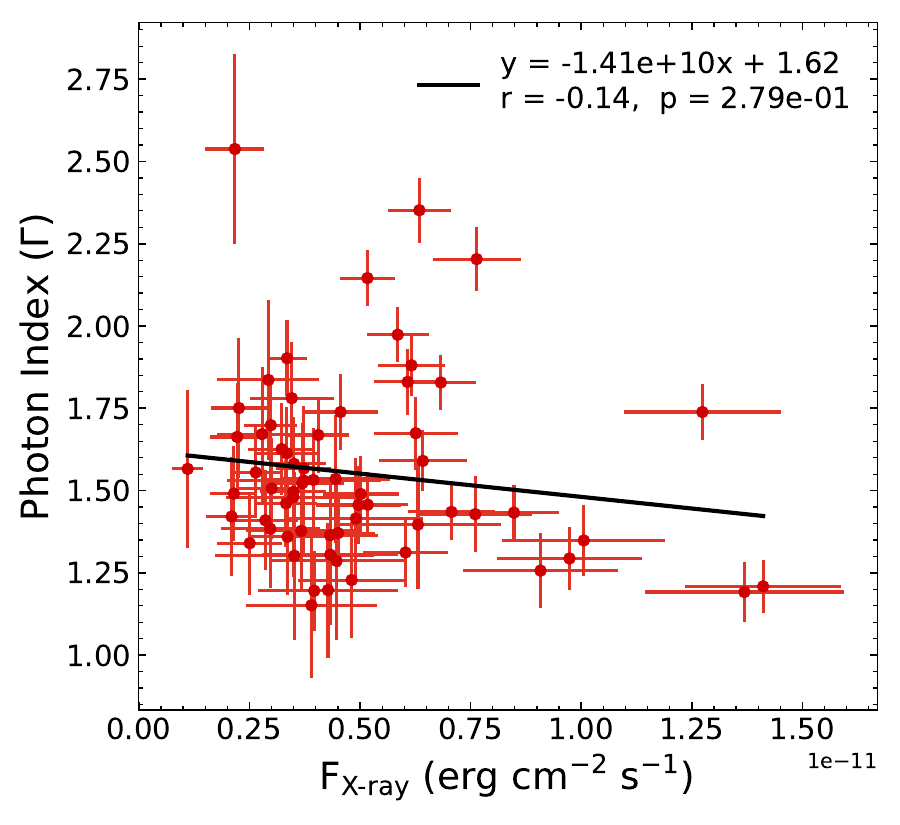}
    \caption{Correlation between X-ray flux and photon index from the absorbed power-law model for all OQ 334 Swift-XRT observations. The black line indicates the linear fit.}
    \label{fig:xrtspec}
\end{figure}

The $\text{AIC}_c$ difference, defined as $\Delta \text{AIC}_{c} = \text{AIC}_{c} - \text{AIC}_{c,\text{min}}$, allows for a comparison of the models with respect to the model that has the lowest $\text{AIC}_c$ value. Following the criteria presented in \citet{Burnham2002}, models with $\Delta \text{AIC}_c > 10$ relative to the best-fit model can be ruled out \citep{Psaradaki2020}. None of the observations showed $\Delta \text{AIC}_c > 10$ for any of the three models, indicating that all models provide statistically acceptable fits. Using the minimum $\text{AIC}_c$ criterion, the power-law model provides the best description for 47 of the 64 observations, the log-parabola model for 13, and the broken power-law model for 4. The Swift-XRT observation log and the best-fitting spectral parameters obtained are presented in Table \ref{tab:xrt_spectra}, and Figure \ref{fig:xrtspec} shows the correlation between flux and photon index from the power-law fit. The photon index varies between $1.15\pm0.22$ and $2.5\pm0.28$ without a systematic dependence on the flux ($r = -0.14,\;p = 0.27$).

\begin{deluxetable*}{lccccccccc}
\tabletypesize{\footnotesize}
\tablecaption{Swift-XRT observation log and best-fitting spectral parameters for OQ 334 in the 0.3--10\,keV band}
\label{tab:xrt_spectra}
\tablehead{
\colhead{ObsID} &
\colhead{Date} &
\colhead{Exposure} &
\multicolumn{2}{c}{powlaw} &
\multicolumn{2}{c}{logpar} &
\multicolumn{3}{c}{bknpow} \\
\cline{4-5} \cline{6-7} \cline{8-10}
\colhead{} &
\colhead{} &
\colhead{(s)} &
\colhead{$\Gamma$} &
\colhead{$F_{\rm X-ray}$} &
\colhead{$\alpha$} &
\colhead{$\beta$} &
\colhead{$\Gamma_1$} &
\colhead{$E_{\rm b}$ (keV)} &
\colhead{$\Gamma_2$} }
\startdata
00010520001 & 2018-01-20 & 1029.6 & $2.54\pm0.29$ & $2.17\pm0.67$ & $2.48\pm0.23$ & $-1.03\pm0.53$ & $2.62\pm0.30$ & $3.96\pm1.49$ & $-2.50^{\dagger}$ \\
00010520002 & 2018-02-22 & 2881.3 & $1.49\pm0.14$ & $2.15\pm0.53$ & $1.29\pm0.23$ & $0.52\pm0.41$ & $1.50\pm0.15$ & $7.12^{\dagger}$ & $-2.46^{\dagger}$ \\
\multicolumn{10}{c}{$\cdots$} \\
00010520072 & 2024-07-31 & 1602.2 & $1.21\pm0.08$ & $14.1\pm1.8$ & $1.21\pm0.14$ & $0.004\pm0.22$ & $1.26\pm0.10$ & $5.00\pm2.14$ & $0.14^{\dagger}$ \\
\enddata
\tablecomments{All models include Galactic absorption fixed at $N_{\rm H}=1.08\times10^{20}$\,cm$^{-2}$, and use \textit{cstat}. \deleted{$F$} \added{$F_{\rm X-ray}$} is the 0.3--10\,keV flux in units of $10^{-12}$\,erg\,cm$^{-2}$\,s$^{-1}$. Parameters marked $^{\dagger}$ had errors that were not constrained by the fit. The full table is available in the supplementary material.}
\end{deluxetable*}

\vspace{-1.5\baselineskip}
\subsubsection{Spectral Hysteresis}
Spectral hysteresis describes the loop-like trajectories traced in the hardness–intensity or flux–photon-index plane during a flare. This phenomenon is driven by the interplay of particle injection, energy-dependent acceleration, and radiative cooling timescales within the jet \citep{Kirk9805, Falcone0401, Tramacere0907}. The direction of this loop (clockwise vs. counter-clockwise) and its geometric morphology serve as a diagnostics for the physical conditions of the emitting region. A clockwise (CW) loop is conventionally associated with a ``soft lag", where the high-energy emission leads the low-energy emission, indicating a regime dominated by synchrotron cooling processes \citep{Takahashi9610, Tramacere0907}. Conversely, a counter-clockwise (CCW) loop is linked to a ``hard lag", which is predicted when the particle acceleration timescale is comparable to or dominates the cooling timescale, causing particles to be gradually accelerated into the radiating window and resulting in low-energy variations leading the high-energy \citep{Falcone0401, Tramacere0907}.

\begin{figure*}
    \centering
    \includegraphics[width=1\linewidth]{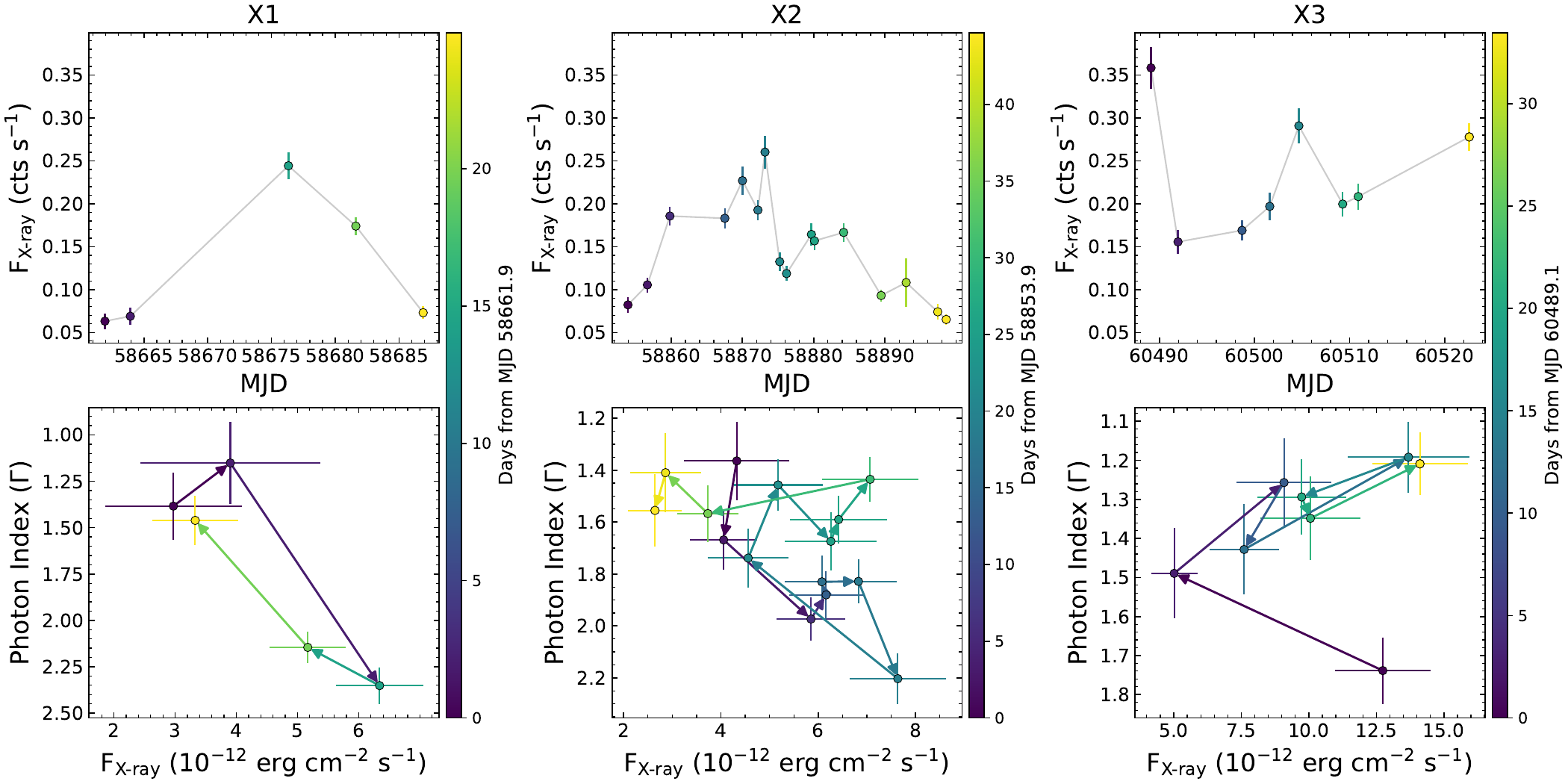}
    \caption{Spectral hysteresis and X-ray light curves of OQ 334 during flares X1, X2, and X3, shown from left to right, respectively. In each column, the upper panel presents the light curve of the corresponding flare, while the lower panel shows its evolution in the flux-photon index plane. Arrows in the lower panels trace the hysteresis loops and denote the direction of temporal evolution. Both panels are colour coded by the time elapsed since the first observation of the flare.}
    \label{fig:hysteresis}
\end{figure*}

To investigate the presence of spectral hysteresis in our source, we examined the relationship between the X-ray flux and the photon index $\Gamma$ for three flares (X1, X2, and X3) identified in the X-ray light curve. These flare periods were selected using the Bayesian Blocks algorithm \citep{Scargle2013}, a statistically robust method for adaptive binning that identifies change points in time series data by segmenting observations into blocks of constant average flux based on a likelihood function. We used the Astropy implementation with a false-alarm probability of $p_0 = 0.05$ to detect different flux states, with flaring periods defined by the threshold $F_{BB} \geq 2 \times \bar{F}$, where $\bar{F}$ is the long-term mean count rate. The identified flares occurred during MJD 58660–58688 (X1), MJD 58852–58900 (X2), and MJD 60485–60524 (X3). The flare light curve and hysteresis plots are shown in Figure \ref{fig:hysteresis}. 

\begin{figure*}
    \centering
    \includegraphics[width=1\linewidth]{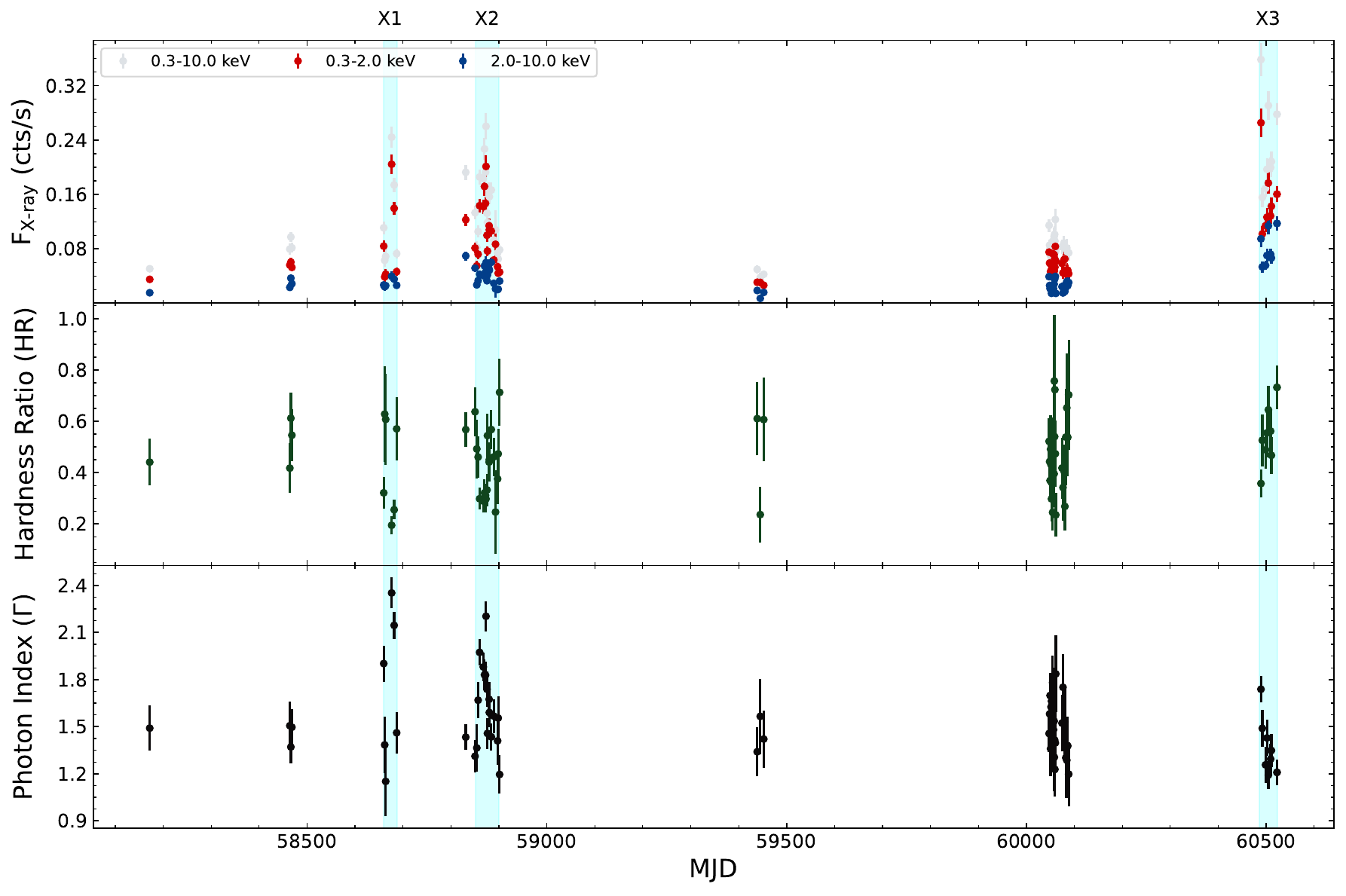}
    \caption{X-ray temporal and spectral variability of OQ 334. From top to bottom, the panels show the light curves in the full (0.3-10.0 keV), hard (2.0-10.0 keV), and soft (0.3-2.0 keV) energy bands; the hardness ratio, defined as the ratio of the hard to soft band count rates; and the photon index obtained from spectral fits. The three flares (X1, X2, and X3) are indicated by shaded regions.}
    \label{fig:xrt_lc_hr_pidx}
\end{figure*}

The X1 flare exhibits a clockwise (CW) loop. The average count rate is $0.125 \pm 0.036~\mathrm{cts\ s^{-1}}$, with the photon index spanning a wide range from $1.151$ to $2.352$. The X2 flare exhibits a complex, multi-loop  trajectory, reflecting multi-peaked light curve morphology. The average count rate during this period is $0.145 \pm 0.014~\mathrm{cts\ s^{-1}}$, with the photon index ranging between $1.364$ and $2.203$. The trajectory initiates with a counter-clockwise (CCW) loop, transitions into a clockwise (CW) loop near the main peak, and reverts back to a CCW loop during the subsequent peak. The first peak in the light curve is spectrally softer, while the second peak is notably harder. The X3 flare is the brightest of the three, with an average count rate of $0.237 \pm 0.022~\mathrm{cts\ s^{-1}}$ and highest brightness of $0.358 \pm 0.024~\mathrm{cts\ s^{-1}}$ at MJD 60489.12. The photon index remained persistently hard throughout the flare, ranging from $1.192$ to $1.739$. Its flux-photon index trajectory is incomplete, as the observations began near the flux peak and did not sample the full rise or decay of the flare. The observed trajectory traces a CW pattern from the high-flux peak, followed by a CCW loop associated with the subsequent peak.

\subsubsection{Hardness Ratio}
To investigate the variability of the X-ray emission, we extracted light curves in three energy bands: soft band (0.3-2.0 keV), hard band (2.0-10.0 keV), and the full  band (0.3-10.0 keV). These light curves are displayed in the first panel of Figure \ref{fig:xrt_lc_hr_pidx}, and the second panel shows the hardness ratio (HR, defined as the ratio of hard to soft counts). The HR provides a model-independent probe of spectral variations across different flux states. The average count rates $0.085 \pm 0.006~\mathrm{cts~s^{-1}}$ in the soft band, $0.038 \pm 0.003~\mathrm{cts~s^{-1}}$ in the hard band, and $0.123 \pm 0.009~\mathrm{cts~s^{-1}}$ in the full band. The soft band dominates the total X-ray flux throughout the observation period. The fractional variability measurements of $57.32 \pm 1.46\%$, $60.29 \pm 2.16\%$, and $55.89 \pm 1.21\%$ for the soft, hard, and full bands respectively, indicate strong variability across all energy ranges, with the hard band showing slightly higher variability than the soft band. 

\begin{deluxetable*}{cccccc}
\tablecaption{X-ray properties of OQ 334 during the three flaring periods.}
\tablehead{
\colhead{Epoch} & 
\colhead{Time Period} & 
\colhead{$\overline{F}_{\mathrm{Soft}}\;(\mathrm{cts\,s^{-1}})$} & 
\colhead{$\overline{F}_{\mathrm{Hard}}\;(\mathrm{cts\,s^{-1}})$} & 
\colhead{HR} & 
\colhead{$\Gamma$} }
\startdata
X1 & $58660.0 - 58688.0$ & $0.094 \pm 0.033$ & $0.030 \pm 0.003$ & $0.323 \pm 0.030$ & $1.97 \pm 0.09$ \\
X2 & $58852.0 - 58900.0$ & $0.105 \pm 0.011$ & $0.039 \pm 0.003$ & $0.374 \pm 0.019$ & $1.71 \pm 0.04$ \\
X3 & $60485.0 - 60524.0$ & $0.153 \pm 0.016$ & $0.085 \pm 0.009$ & $0.554 \pm 0.027$ & $1.34 \pm 0.06$ \\
\enddata
\tablecomments{X1, X2, and X3 denote the three X-ray flares, and the time periods are given in MJD. $\overline{F}_{\mathrm{Soft}}$ and $\overline{F}_{\mathrm{Hard}}$ are the mean count rates in the soft (0.3-2.0 keV) and hard (2.0-10.0 keV) bands, respectively. HR is the hardness ratio, defined as the ratio of the hard to soft band count rates, and $\Gamma$ is the photon index from the absorbed power-law fits.}
\label{tab:flare_xray}
\end{deluxetable*}

\begin{figure}
    \centering
    \includegraphics[width=1\linewidth]{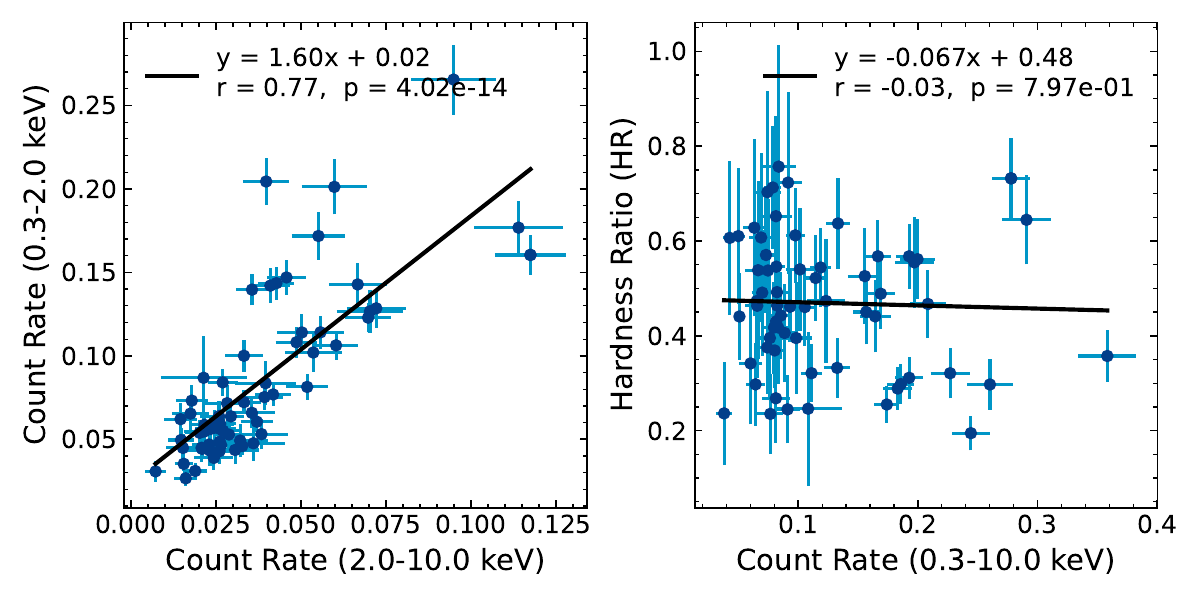}
    \caption{Correlation plots from the Swift-XRT hardness ratio (HR) analysis of OQ 334. The left panel shows the hard-band (2.0–10.0 keV) versus soft-band (0.3–2.0 keV) count rates, while the right panel shows the full-band (0.3–10.0 keV) count rate versus HR. The black lines indicate the corresponding linear fits.}
    \label{fig:xray_hr}
\end{figure}

\vspace{-2.0\baselineskip}
Figure \ref{fig:xray_hr} presents the correlations between energy bands and HR in two panels. A strong positive correlation is found between the hard and soft-band count rates (Pearson's $r = 0.77$, $p = 4.02 \times 10^{-14}$), while no significant correlation is found between the full-band count rate and HR (Pearson's $r = -0.03$, $p = 7.97 \times 10^{-1}$), consistent with the absence of any trend between flux and photon index. Table \ref{tab:flare_xray} quantifies the spectral evolution across the three flares. The first two flares (X1 and X2) exhibit relatively similar spectral characteristics, with soft-band-dominated emission (HR $\sim 0.32$-$0.37$) and photon indices ($\Gamma = 1.97 \pm 0.09$ and $1.71 \pm 0.04$). In contrast, the third flare (X3) displays different behaviour. While it reaches the highest flux level, it is significantly harder (HR $= 0.554 \pm 0.027$, $\Gamma = 1.34 \pm 0.06$). 

\subsection{Changing-look Nature}
\begin{figure*}
    \centering
    \includegraphics[width=1\linewidth]{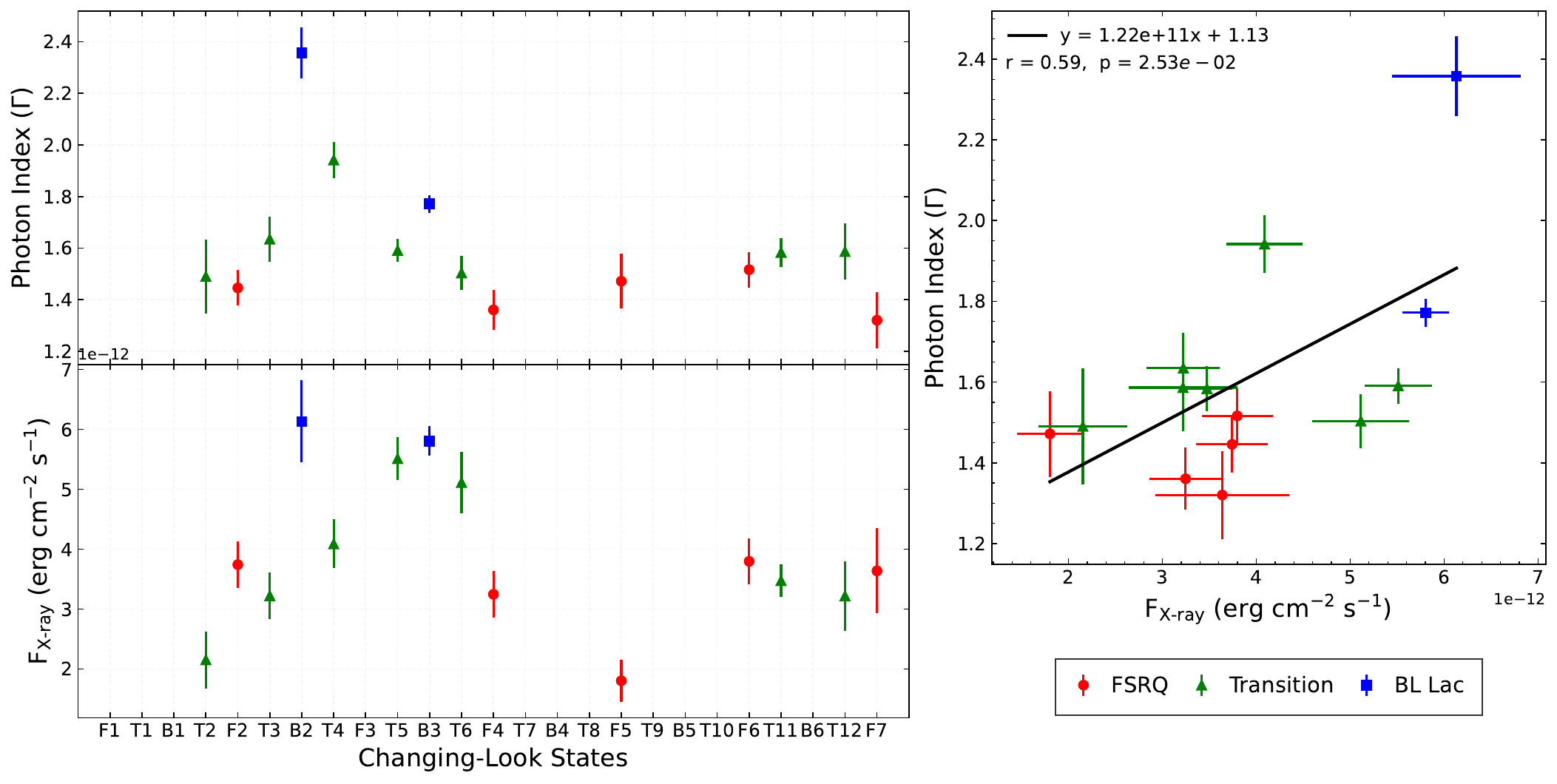}
    \caption{Swift-XRT analysis of the changing-look (CL) states of OQ 334 identified by \citet{Mishra2021} and \citet{Ren2024a}. The left panel shows the photon index (top) and the X-ray flux (bottom) for each CL state, while the right panel shows the correlation between the flux and photon index during these states. The black line indicates the corresponding linear fit.}
    \label{fig:xray_clb}
\end{figure*}

To investigate the changing-look (CL) behaviour of the source, we adopted the state classification from Table 1 of \citet{Ren2024a} and combined the Swift-XRT observations corresponding to each CL states. The spectral analysis shown in Figure \ref{fig:xray_clb} reveals a clear and systematic separation of the three spectral states, BL Lac, FSRQ, and Transition, in the X-ray flux-photon index space. This evolution of the X-ray photon index ($\Gamma_X$) closely mirrors the behaviour of the $\gamma$-ray photon index ($\Gamma_\gamma$) during CL states quantified by \citet{Ren2024a}. In their work, CL states are defined via the combined evolution of broad-line EW and $\Gamma_\gamma$, with characteristic ranges: $\Gamma_\gamma \gtrsim 2.2$ for FSRQ states, $2.0 < \Gamma_\gamma < 2.2$ for transition states, and $\Gamma_\gamma \lesssim 2.0$ for BL Lac states. The X-ray spectral evolution follows an opposite trend to that observed in $\gamma$ rays: FSRQ states exhibit the hardest X-ray spectra, with $\Gamma_X$ clustered at $\Gamma_X \approx 1.32-1.52$, while BL Lac states show softer X-ray spectra, with $\Gamma_X \approx 1.77-2.36$. Transition states occupy an intermediate space with $\Gamma_X \approx 1.49-1.94$. This clustering indicates that the X-ray emission is coupled to the CL state of the blazar. Within CL states, we find a positive correlation between $F_{\text{X-ray}}$ and $\Gamma_X$ ($r = 0.59$, $p = 2.53 \times 10^{-2}$), indicating a ``softer-when-brighter'' trend in the X-rays.

\begin{figure*}
    \centering
    \includegraphics[width=1\linewidth]{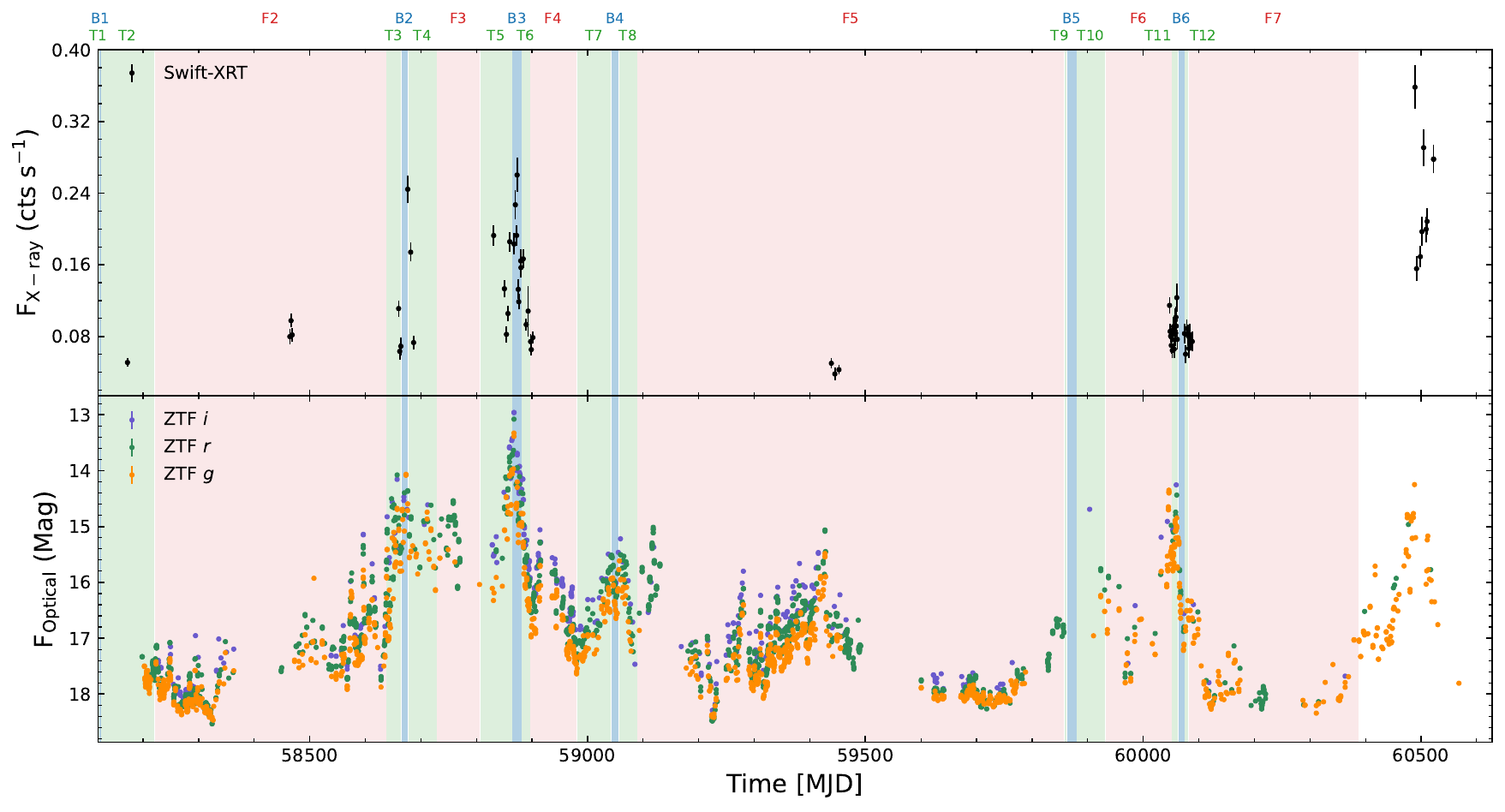}
    \caption{Swift-XRT and ZTF light curves of the OQ 334 changing-look (CL) states of OQ 334 identified by \citet{Mishra2021} and \citet{Ren2024a}. Shaded regions indicate CL states: BL Lac (blue), transition (green), and FSRQ (red).}
    \label{fig:lc_clb}
\end{figure*}

The X-ray flares X1 and X2 coincide with the CL states (Figure \ref{fig:lc_clb}). Flare X1 overlaps the BL Lac state B2 and the transition states T3 and T4, while flare X2 spans T5, B3, T6, and F4. Flare X3 occurred after the time period covered by the previous CL studies. In both flares X1 and X2, $\Gamma_X$ increases from the transition state into the BL Lac state and then decreases again in the subsequent transition state, tracing a rise-and-fall pattern across the CL sequence. Using long-term ZTF photometry, \citet{Chand2025} showed that FSRQ $\rightarrow$ BL Lac transitions were accompanied by an optical flare of $\approx 1.5-2.5$ mag. For Flare X1, the optical counterpart has amplitudes of $\Delta m_g = 1.734$ mag, $\Delta m_r = 1.119$ mag, and $\Delta m_i = 0.939$ mag. Flare X2 shows exceptional optical variability by $\Delta m_g = 3.651$ mag, $\Delta m_r = 3.465$ mag, and $\Delta m_i = 2.983$ mag. Flare X3 flare with $\Delta m_g = 2.097$ mag, no $r$- or $i$-band data are available.

\subsection{Broadband SED} \label{sec:sed}
The broadband SEDs of X-ray flares were modelled with the standard one-zone leptonic framework using the open-source Python package \texttt{agnpy} (v0.5.1, \citealt{Nigro2204}). The model assumes a single homogeneous spherical emission region of radius $R_{\rm b}$, filled isotropically with a uniform magnetic field $B$ and a population of relativistic electrons. The blob moves along the jet with bulk Lorentz factor $\Gamma$ and velocity $\beta c$, with the observed emission Doppler boosted by $\delta_{\rm D} = [\Gamma(1 - \beta\cos\theta)]^{-1}$, where $\theta$ is the viewing angle. The size of the emission region is further constrained by the shortest observed variability timescale $t_{\rm var}$ through the causality condition $R_{\rm b} \lesssim c\,\delta_D\,t_{\rm var}/(1 + z)$.  For OQ 334, the shortest $\gamma$-ray variability timescale reported is $\sim9$ hours \citep{Prince2021}. The relativistic electron population is described by a broken power-law energy distribution,
\begin{equation}
\begin{split}
n(\gamma) = k \Bigg[ & \left(\frac{\gamma}{\gamma_b}\right)^{-p_1} H(\gamma; \gamma_{\rm min}, \gamma_b) \\
& + \left(\frac{\gamma}{\gamma_b}\right)^{-p_2} H(\gamma; \gamma_b, \gamma_{\rm max}) \Bigg],
\end{split}
\end{equation}
where $\gamma$ is the electron Lorentz factor in the comoving frame, $k$ is the normalisation, and $p_1$ and $p_2$ are the indices below and above the break ($\gamma_b$). The Heaviside functions ($H$) restrict each power-law component to its respective energy range.

The broadband emission is reproduced by three non-thermal radiative components: synchrotron radiation, synchrotron self-Compton (SSC) emission, and external Compton (EC) scattering from the broad-line region (BLR) and dusty torus (DT), along with thermal components from the accretion disk and dusty torus. The low-energy hump of the SED is modelled as synchrotron emission from relativistic electrons in the emitting blob, accounted for synchrotron self-absorption. The same electron population upscatters both the synchrotron photons via IC scattering, producing the SSC component, and photons from the external radiation fields, producing the EC components responsible for the high-energy hump of the SED. The SED of each component is calculated using the equations given in Appendix A of \citet{Nigro2204}.

The luminosities and characteristic radii of the two external fields scale with the accretion disk luminosity ($L_{\rm disk}$) following \citealt{Ghisellini0908}. The BLR radius is given by $R_{\rm BLR} = 10^{17} L_{\rm d,45}^{1/2}\ {\rm cm}$, where $L_{\rm d,45}=L_{\rm disk}/(10^{45}\ {\rm erg\ s^{-1}})$, while the DT radius is $R_{\rm DT} = 2.5\times10^{18} L_{\rm d,45}^{1/2}\ {\rm cm}$. The BLR is modelled as an infinitesimally thin spherical shell of radius $R_{\rm BLR}$ reprocessing a fraction $\xi_{\rm line}$ of the disk radiation and re-emitting it as a monochromatic Lyman-$\alpha$ line \citep{Finke1610}. The DT is modelled as an infinitesimally thin ring of radius $R_{\rm DT}$ lying in the plane perpendicular to the jet axis, which reprocesses a fraction $\xi_{\rm DT}$ of the disk radiation and re-emits it as thermal infrared radiation characterised by a single blackbody temperature $T_{\rm DT}$ \citep{Finke1610}. The disk luminosity is fixed to $L_{\rm disk} = 2 \times 10^{46}$ erg s$^{-1}$, estimated from the broad Mg\,\textsc{ii} emission line flux measured during the 2020 multi-wavelength campaign \citep{MAGICCollaboration2021}. The corresponding radii of the two external fields are $R_{\rm BLR} = 4.4 \times 10^{17}$~cm and $R_{\rm DT} = 1.1 \times 10^{19}$~cm. The reprocessing fractions are set to $\xi_{\rm line} = 0.1$ for the BLR and $\xi_{\rm DT} = 0.6$ for the DT, and the torus dust is modelled as a blackbody at $T_{\rm DT} = 1000$ K.

\begin{figure}
    \centering
    \includegraphics[width=1\linewidth]{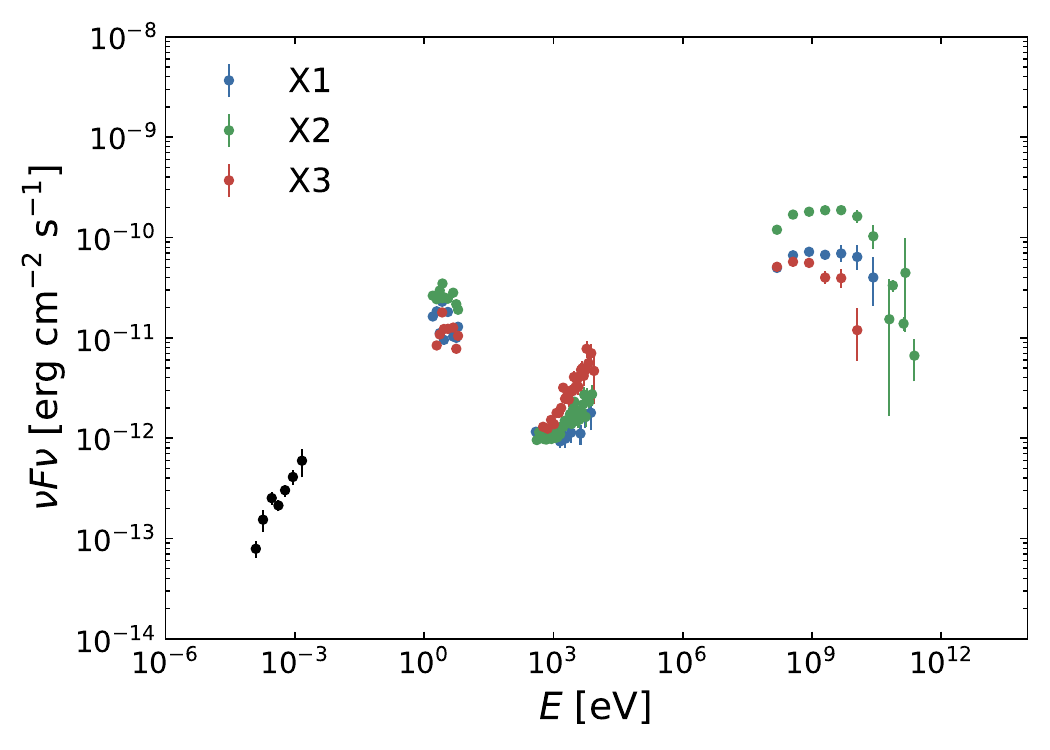}
    \caption{Broad-band SEDs of OQ 334 during the three X-ray flaring periods X1, X2, and X3, shown in blue, green, and red, respectively. The Planck data points, plotted in black, are common to all three flares.}
    \label{fig:sed_epochs}
\end{figure}

The SED data are supplemented with non-simultaneous Planck radio measurements retrieved from the Markarian Multi-wavelength Data Center \citep{Sahakyan2412, PlanckCollaboration1811}. Because the high-energy emission region in OQ 334 is physically offset from the radio-emitting core by $\approx 11$ pc, the observed radio emission is dominated by the slowly varying larger-scale jet \citep{Prince2021}, and these measurements are therefore used only to define the low-frequency part of the synchrotron component. The SED of the flares are shown in Figure \ref{fig:sed_epochs}. The radiative processes are combined into a single spectral model through the \texttt{ExternalComptonModel} class of \texttt{agnpy}, which interfaces with the \texttt{modeling} sub-package of \texttt{gammapy} (\citealt{Donath2310}). We first obtain a best-fit solution by minimising the $\chi^{2}$ statistic with the \texttt{iminuit} optimizer (\citealt{iminuit}), and then use the resulting parameter values to initialise a Markov Chain Monte Carlo (MCMC) sampling performed with \texttt{emcee} (\citealt{ForemanMackey1303}). Uniform priors are assigned to all free model parameters, and the resulting posterior probability distributions are used to derive the parameter values and their uncertainties.

The electron Lorentz factors are held fixed throughout, at $\gamma_{\rm min} = 10$, following \citet{MAGICCollaboration2021}, which showed that such a value is necessary to correctly reproduce the valley observed between the synchrotron and IC humps, and at $\gamma_{\rm max} = 1 \times 10^{7}$. The location of the emission region along the jet axis was left free in the initial MCMC sampling. The sampling consistently converged to $r \approx 1.5 \times 10^{19}$~cm, far outside the BLR boundary and beyond the characteristic radius of the DT, where the BLR contribution is negligible, and the EC scattering on DT becomes the dominant high-energy process \citep{MAGICCollaboration2021, Peer2026Jul}. The value of $r$ is therefore fixed in the final sampling, leaving six free parameters: $k_{\rm e}$, $p_1$, $p_2$, $\gamma_b$, $\delta_{\rm D}$, $B$.

\begin{deluxetable*}{ccccccc}
\tablecaption{Fitted SED model parameters of OQ 334 during the X-ray flares}
\label{tab:sed_params}
\tablehead{
\colhead{Epoch} &
\colhead{$k_{\rm e}$} &
\colhead{$p_1$} &
\colhead{$p_2$} &
\colhead{$\gamma_{\rm b}$} &
\colhead{$\delta_{\rm D}$} &
\colhead{$B$} }
\startdata
X1 & $7.09^{+3.39}_{-2.46} \times 10^{-4}$ & $1.95^{+0.08}_{-0.08}$ & $4.05^{+0.08}_{-0.06}$ & $5.24^{+0.84}_{-0.68} \times 10^{3}$ & $34.20^{+1.43}_{-1.37}$ & $0.136^{+0.008}_{-0.007}$ \\
X2 & $3.33^{+1.13}_{-0.88} \times 10^{-4}$ & $1.94^{+0.04}_{-0.04}$ & $4.75^{+0.12}_{-0.10}$ & $9.23^{+1.11}_{-0.91} \times 10^{3}$ & $34.69^{+0.96}_{-1.01}$ & $0.113^{+0.004}_{-0.004}$ \\
X3 & $5.99^{+3.21}_{-2.06} \times 10^{-4}$ & $2.20^{+0.05}_{-0.06}$ & $4.87^{+0.10}_{-0.18}$ & $8.15^{+0.96}_{-0.97} \times 10^{3}$ & $28.01^{+1.18}_{-1.21}$ & $0.136^{+0.009}_{-0.008}$ \\
\enddata
\tablecomments{Quoted values are posterior medians with $16$th and $84$th percentile uncertainties. $k_{\rm e}$ is the electron distribution normalization (cm$^{-3}$); $p_1$ and $p_2$ are the low- and high-energy electron spectral indices; $\gamma_{\rm b}$ is the break Lorentz factor; $\delta_{\rm D}$ is the Doppler factor; and $B$ is the magnetic field strength (G).}
\end{deluxetable*}

\begin{deluxetable*}{cccccccc}
\tablecaption{Derived SED properties of OQ 334 during the X-ray flares}
\label{tab:sed_derived}
\tablehead{
\colhead{Epoch} &
\colhead{$\nu_{\rm syn}^{\rm peak}$} &
\colhead{$\nu_{\rm IC}^{\rm peak}$} &
\colhead{CD} &
\colhead{$U_{\rm e}$} &
\colhead{$U_{\rm B}$} &
\colhead{$U_{\rm e}/U_{\rm B}$} &
\colhead{$P_{\rm jet}$} \\
\colhead{} &
\colhead{(Hz)} &
\colhead{(Hz)} &
\colhead{} &
\colhead{(erg cm$^{-3}$)} &
\colhead{(erg cm$^{-3}$)} &
\colhead{} &
\colhead{(erg s$^{-1}$)} }
\startdata
X1 & $2.92 \times 10^{14}$ & $3.08 \times 10^{23}$ & $6.07$ & $9.24 \times 10^{-2}$ & $7.31 \times 10^{-4}$ & $126$ & $1.05 \times 10^{47}$ \\
X2 & $6.06 \times 10^{14}$ & $8.57 \times 10^{23}$ & $9.28$ & $1.38 \times 10^{-1}$ & $5.04 \times 10^{-4}$ & $274$ & $1.48 \times 10^{47}$ \\
X3 & $3.91 \times 10^{14}$ & $6.16 \times 10^{22}$ & $5.57$ & $4.63 \times 10^{-1}$ & $7.35 \times 10^{-4}$ & $631$ & $5.87 \times 10^{47}$ \\
\enddata
\tablecomments{$\nu_{\rm syn}^{\rm peak}$ and $\nu_{\rm IC}^{\rm peak}$ are the synchrotron and inverse-Compton peak frequencies of the best-fit model; CD is the Compton dominance, defined as the ratio of the inverse-Compton to synchrotron peak $\nu F_\nu$. $U_{\rm e}$ and $U_{\rm B}$ are the comoving electron and magnetic energy densities, and $U_{\rm e}/U_{\rm B}$ is the departure from equipartition. $P_{\rm jet}$ is the total jet power, summing the electron, cold-proton and magnetic contributions, computed assuming one cold proton per emitting electron.}
\end{deluxetable*}

\begin{figure}
    \centering
    \subfigure{\includegraphics[width=0.5\textwidth]{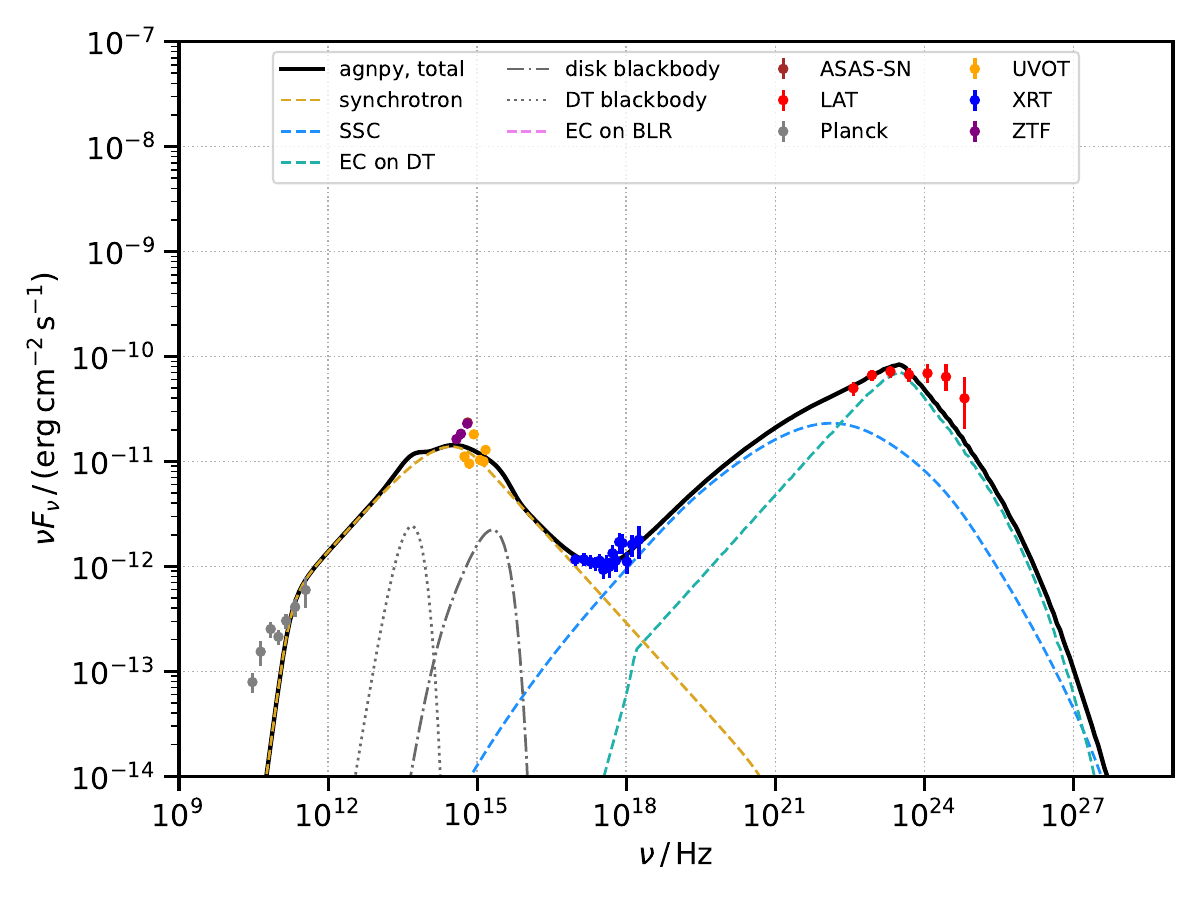}}
    \subfigure{\includegraphics[width=0.5\textwidth]{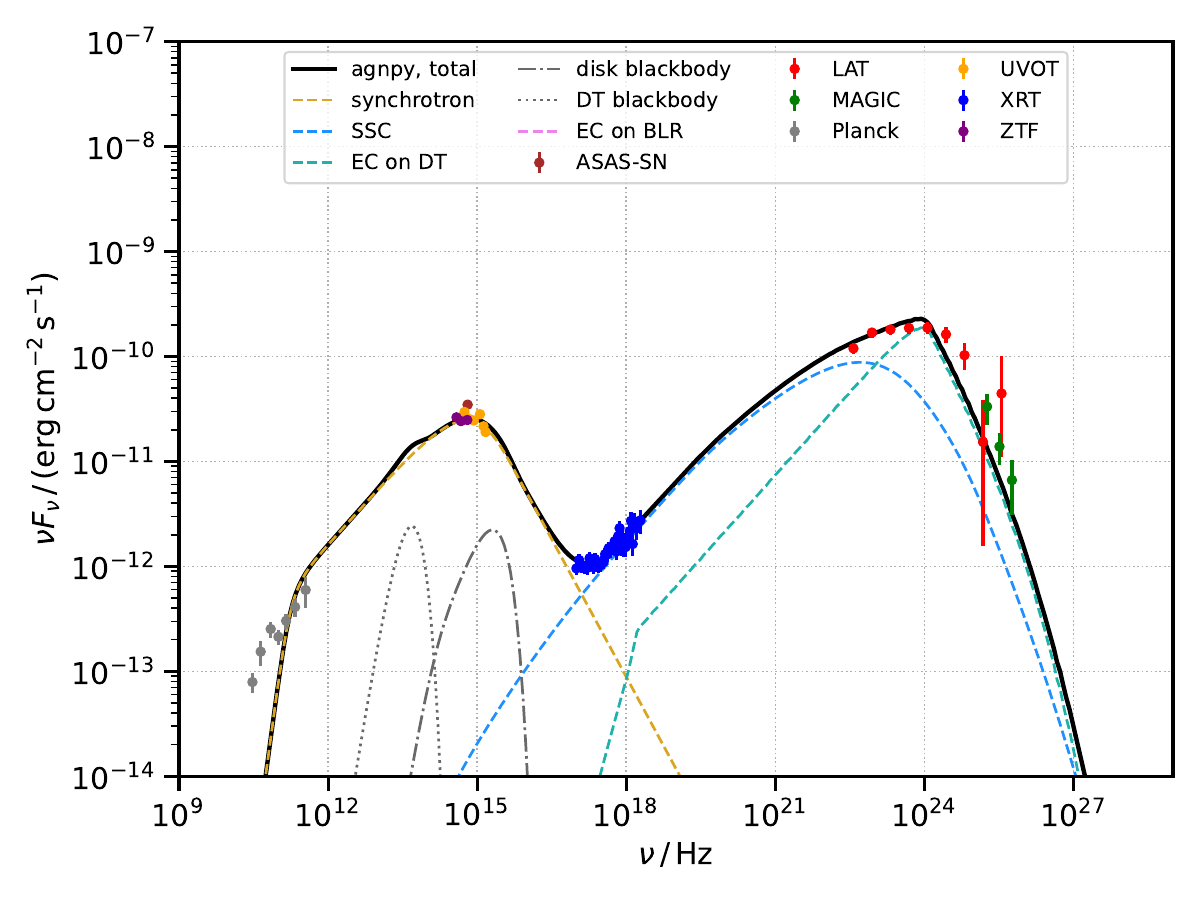}}
    \subfigure{\includegraphics[width=0.5\textwidth]{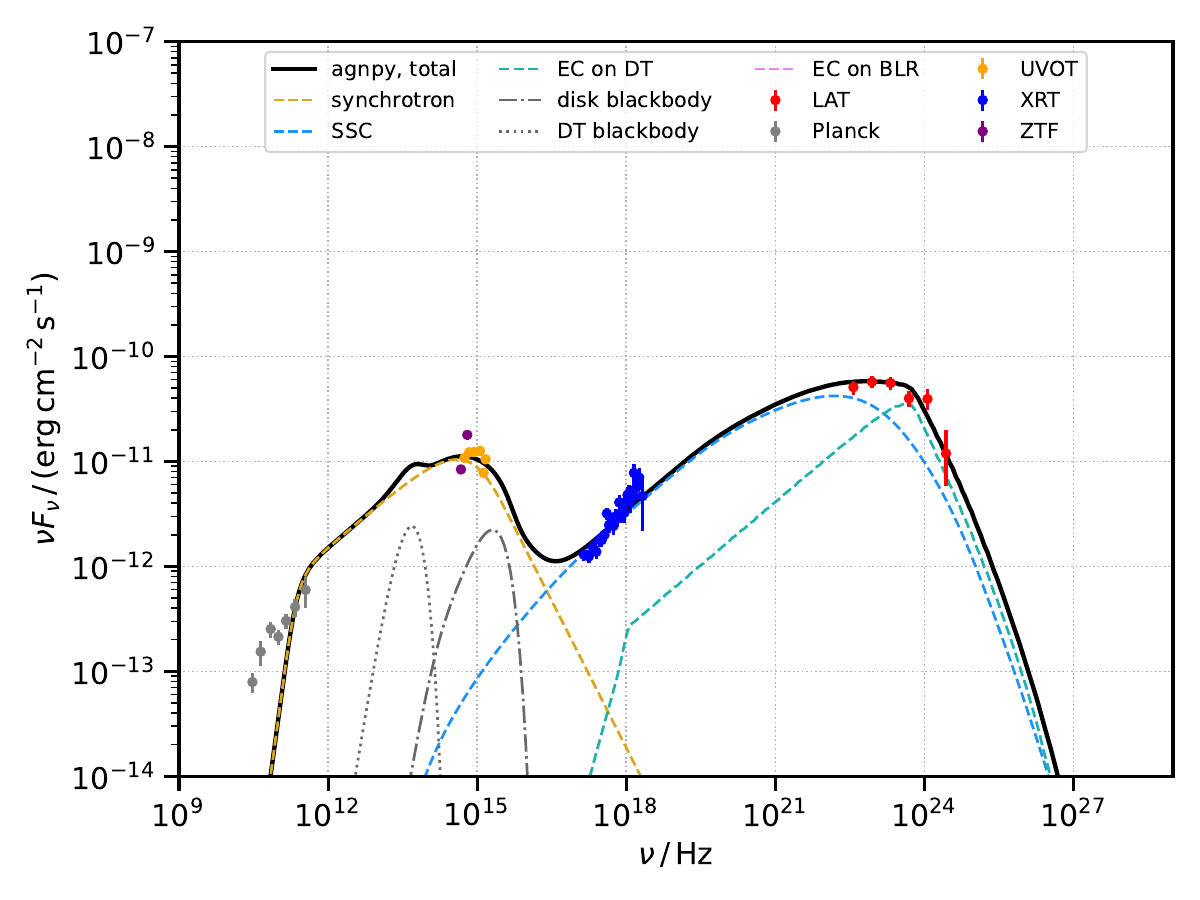}}
    \caption{Broadband SED modelling of OQ 334 during the three X-ray flares X1, X2, and X3, shown from top to bottom, respectively. In each panel, the individual radiative components and the observed data points are identified in the legend.}
    \label{fig:sed_fit}
\end{figure}

\vspace{-4.0\baselineskip}
The broadband SED and the best-fit model are shown in Figure \ref{fig:sed_fit}. Table \ref{tab:sed_params} reports the fitted values of the free parameters, and Table \ref{tab:sed_derived} reports the derived SED quantities. The break Lorentz factor rises from X1 to X2 and stays at a comparable level at X3, whereas the electron normalisation shows no clear flare-to-flare trend. The low-energy index is almost the same between X1 and X2 but softens significantly at X3, while the high-energy index increases from X1 to X2 and then remains at this higher value at X3. As a result, the spectral break $p_2-p_1$ increases from X1 to the later epochs. The Doppler factor is consistent between X1 and X2 and drops at X3. The magnetic field remains low throughout, being identical within errors at X1 and X3 and decreasing slightly at X2. The synchrotron peak frequency varies by a factor of $\sim2$ across the three epochs, while the IC peak shifts by more than an order of magnitude, reaching its highest value at X2 and dropping at X3. Compton dominance follows a similar pattern, peaking at X2 and declining at X3. The energy densities and jet powers indicate a particle-dominated emission region that departs strongly from equipartition in all three flaring epochs. The magnetic energy density remains relatively stable, whereas the electron energy density increases progressively from X1 to X3. Consequently, the departure from equipartition strengthens across the flares, with $U_{\rm e}/U_{\rm B}$ rising from X1 to X3. This increase in particle density is reflected in the total jet power, computed as $P_{\rm jet} = 2 \pi R_b^2 \Gamma^2 \beta c U_{\rm i},$ where $U_{\rm i}$ represents the energy density of the components: electrons, cold protons, and magnetic field. The resulting jet power grows from X1 to X3, assuming one cold proton per relativistic electron \citep{Celotti0803}.

\section{Discussion} \label{sec:discussion}

\subsection{Timing \& Optical Analysis}
The cross-correlation analysis provides evidence for a close temporal connection between the inter-band emission. The near-zero lag suggests that the optical and $\gamma$-ray emitting components respond to the same underlying variations in the relativistic jet. Given the weekly binning of the LAT light curve, this lag indicates that the optical and $\gamma$-ray emission originate in the same region. A radio–$\gamma$-ray connection is also present in this source, with radio variations leading the $\gamma$-ray emission by $\sim 70$ days \citep{Prince2021}. The redder-when-brighter behaviour observed is commonly reported in FSRQs and is generally attributed to the increasing dominance of the variable synchrotron emission from the jet, which has a redder spectrum, over the comparatively stable thermal emission from the accretion disc as the source brightens \citep{Chand2025}. The correlation between the two colour indices shows that the colour changes are consistent across the optical bands, in line with the colour–magnitude diagrams.

\subsection{X-ray Analysis}
The $\text{AIC}_c$-based model comparison indicates that, although all three spectral models provide statistically acceptable descriptions of the X-ray spectra, the majority of the observations are best represented by a simple power-law shape. This suggests that a single-component spectrum suffices to describe the X-ray emission of the source in most observations, with no strong requirement for additional curvature or spectral breaks. The absence of a clear correlation between the X-ray flux and the photon index implies that variations in the source brightness are not accompanied by systematic spectral hardening or softening over long timescales. In BL Lacs, the photon index generally shows a dependence on flux, with a harder-when-brighter trend when the X-ray band samples the synchrotron tail and a softer-when-brighter trend when the synchrotron and IC components overlap in the band \citep{Giommi2021Nov, Williamson2014Jul}. FSRQs, in which the X-ray emission is generally dominated by the IC component, show a harder-when-brighter trend \citep{Gu2011Jun, Hayashida2012Aug, Sahakyan2021Jul} or no correlation over long timescales \citep{DAmmando2019Dec, Zacharias2026Aug}. The absence of a long-term correlation in the source is closer to the behaviour reported for FSRQs than to that of BL Lacs.

The distinct morphologies and directions of the hysteresis loops observed across the three flaring periods reveal a dynamic emitting environment with shifting physical regimes. During Flare X1, the CW loop in the flux-$\Gamma$ plane is consistent with energy-dependent radiative cooling, in which the highest-energy electrons cool more rapidly and spectral softening follows the flux decline \citep{Kirk9805, Falcone0401}. Flares X2 and X3 display complex, multi-loop trajectories that transition between CCW and CW directions, signifying a departure from the simple cooling-dominated state. For Flare X3, the hard photon index ($\Gamma$ ranging from 1.192 to 1.739) suggests that the high-energy electron population was continuously replenished by rapid particle acceleration or sustained injection, preventing the cooling-induced softening seen in X1.

\begin{figure}
    \centering
    \includegraphics[width=0.9\linewidth]{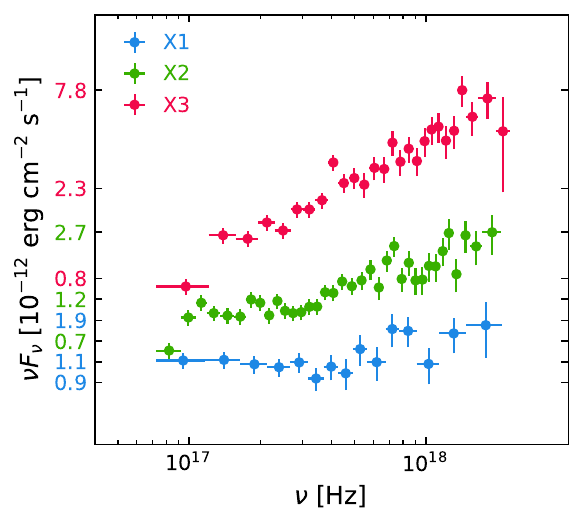}
    \caption{Swift-XRT spectra of OQ 334 during the three X-ray flares X1, X2, and X3, fitted with an absorbed power-law model. Vertical offsets have been applied to the spectra for clarity of presentation, and each spectrum is accompanied by a colour-matched vertical axis.}
    \label{fig:xrt_flare}
\end{figure}

The hardness ratio analysis provides a robust, direct probe of spectral variability that complements the spectral fitting. The strong positive correlation between the hard-band and soft-band count rates shows that flux variations across the X-ray spectrum are tightly coupled. Three flaring periods reveal a progressive, monotonic hardening of the X-ray band. The photon index decreases systematically across the three flares, $\Gamma = 1.97 \pm 0.09$ (X1) $\rightarrow 1.71 \pm 0.04$ (X2) $\rightarrow 1.34 \pm 0.06$ (X3), while the mean HR increases in lock-step, $\mathrm{HR} = 0.323 \pm 0.030 \rightarrow 0.374 \pm 0.019 \rightarrow 0.554 \pm 0.027$. This monotonic evolution implies that the spectral changes are not purely stochastic but trace a physical transition in the X-ray flares. The progression from a soft flaring state to a hard flaring state indicates that (Figure \ref{fig:xrt_flare}), as the source becomes increasingly populated by high energy particles, the X-ray emitting region undergoes a systematic change in its underlying physical conditions. 

\subsection{Changing-look Nature}
The flux-$\Gamma_X$ analysis of the CL states shows that the X-ray spectral state of the source is directly coupled to its CL classification. The measured photon index reflects the relative contribution of the synchrotron and IC components to the X-ray band. The hard spectra of the FSRQ states ($\Gamma_X \lesssim 1.5$) are consistent with IC-dominated emission, whereas the soft spectra of the BL Lac states ($\Gamma_X \gtrsim 1.8$) indicate an increased contribution of the synchrotron tail to the soft X-ray band, with the transition states occupying an intermediate regime. The ``softer-when-brighter'' trend within the CL states confirms this interpretation, implying that the flux enhancements during these states are driven by the synchrotron component, whose steep spectrum progressively softens the composite emission as it brightens.

Flares X1 and X2 includes the BL Lac states B2 and B3 from the CL classification (Figure \ref{fig:xray_clb}). The BL Lac states correspond to phases of elevated jet activity rather than a quiescence. The optical flares reported by \citet{Chand2025} coincide with the flares X1 and X2, thereby extending the flare-CL transition association to the X-ray band. This association can be examined only for B2 and B3, as X-ray observations are not available for the other BL Lac states. This enhanced jet continuum can dilute the broad Mg$_{\rm II}$ line below the EW threshold defining the BL Lac state. Although simultaneous optical spectroscopy is needed to confirm this scenario directly, the optical and X-ray behaviour during B2/X1 and B3/X2 is consistent with jet-continuum dilution of the BELs rather than an intrinsic change in the accretion disk \citep{Mishra2021, Chand2025}. This is consistent with spectroscopic studies of other CLB sources, which show strong variability in the optical continuum and high-energy emission while the BELs remain comparatively stable \citep{Chen2512, Azzollini2511, Pandey2501}. Flare X3 challenges the interpretation that the CL transitions trace a systematic evolution of the central engine \citep{Ren2024, Ren2024a}. If the association between X-ray flaring and spectral softening observed during the CL states extended to the full observation period, the brightest outburst would be expected to display the most pronounced softening. Instead, Flare X3, has the hardest photon index of the three flares ($\Gamma_X = 1.34 \pm 0.06$), characteristic of the FSRQ states ($\Gamma_X \lesssim 1.5$). The persistence of a hard spectrum at peak brightness indicates that the flux enhancement during X3 is driven by the IC component.

\subsection{Broadband SED}
The one-zone leptonic model successfully reproduces the broadband emission across all three flares, with external seed photons supplied by the thermal radiation of the dusty torus. The X-ray emission is produced by the SSC component, which is strong and extends towards the IC peak, whereas the $\gamma$-ray emission is dominated by EC-DT scattering. SED modelling results are plotted in Figure \ref{fig:sed_fit}. The magnetic field remains weak and approximately constant ($B \approx 0.11\text{--}0.14$), indicating that the flaring activity is driven by changes in the particle population rather than magnetic-field variations. The rise in $P_{\rm jet}$ and in $U_{\rm e}/U_{\rm B}$ at X3 is due to the increase in $U_{\rm e}$ from the injection of kinetic energy into the emitting electron population, consistent with the particle-dominated flaring scenario \citep{Prince2021}. The emission region remains far from equipartition in all three epochs, with $U_{\rm e}/U_{\rm B} \gg 1$, so a magnetically dominated region is not required to reproduce the broadband SEDs. For the VHE flare, which falls within X2, \citet{MAGICCollaboration2021} located the emission region inside the DT ring, which yields a higher magnetic field and an emission region close to equipartition. The jet power is high in all three flares and exceeds $L_{\rm edd}$. The power carried by FSRQ jets is often larger than the disc luminosity and can surpass the Eddington luminosity \citep{Ghisellini0911}.

The synchrotron peak frequencies derived for X1, X2, and X3 all lie within the intermediate-synchrotron-peaked (ISP) band under the standard definition $\nu_{\rm syn}^{\rm peak}=10^{14}$–$10^{15}\,\mathrm{Hz}$, indicating that the synchrotron component does not migrate along the blazar sequence across these three flares. Although the source is classified as LSP in the 4LAC catalogue on the basis of a time-averaged synchrotron peak at $3.16\times10^{13}\,\mathrm{Hz}$, our measured peak frequencies are an order of magnitude higher and are consistent with the ISP classification of \citet{Prince2021}, whereas \citet{Ren2024, Ren2024a} place the peak below $10^{14}\,\mathrm{Hz}$. The high-energy peak shows a different behaviour, with the IC peak displaced well above the long-term average during X1 and X2 but returning close to it during X3.

For an ISP blazar like OQ 334, the Swift-XRT band ($\approx 10^{17}-10^{18}$ Hz) lies in the spectral valley between the declining high-energy tail of the synchrotron hump and the rising branch of the IC component. In this regime, the observed X-ray spectrum is a superposition of synchrotron and IC emission, and its shape is sensitive to their relative contributions. The three X-ray flares reveal a clear spectral evolution that reflects a changing balance between these two components. Flare X1 shows the softest X-ray spectrum, with SED model indicating a substantial synchrotron contribution in addition to the IC component. During X2, the spectrum hardens relative to X1, with a reduced synchrotron fraction and a correspondingly larger IC contribution. By X3, the spectrum is the hardest of the three, and the SED modelling shows that the X-ray emission is dominated entirely by the IC component.

\section{Conclusion} \label{sec:conclusions}
We have presented a multi-wavelength study of the CL TeV blazar OQ 334 (B2 1420+326) spanning MJD 54769.0 to 60824.0, focusing on the X-ray regime. The analysis centres on the CL nature of the source and the X-ray flaring activity identified over this interval. The main results are summarised as follows.

\begin{enumerate}
    \item The source exhibits intense, high-amplitude non-thermal variability across all observed wavebands, with tight optical, X-ray, and $\gamma$-ray coupling, while the optical emission shows significant colour-colour and colour-magnitude correlations.
    \item The $\text{AIC}_c$ comparison shows that all three spectral models provide statistically acceptable fits, with a simple power law favoured for the majority of Swift-XRT observations and log-parabolic or broken power-law models preferred in fewer cases.
    \item Spectral hysteresis analysis reveals distinct evolution across the three flares: X1 shows a clockwise loop consistent with cooling-dominated emission, while X2 and X3 exhibit complex multi-loop trajectories with changing directions, indicating shifts in the dominant physical processes.
    \item The hardness ratio analysis reveals a systematic hardening across the three X-ray flares, with the photon index decreasing from $\Gamma = 1.97 \pm 0.09$ (X1) to $1.71 \pm 0.04$ (X2) and $1.34 \pm 0.06$ (X3), indicating a progressive change in the physical conditions of the X-ray flaring.
    \item The X-ray spectra systematically distinguish the CL states, with FSRQ states being harder, BL Lac states softer, and transition states intermediate. The CL behaviour is therefore consistent with jet-driven dilution of the BELs by a variable synchrotron continuum.
    \item The broadband SEDs are well described by a one-zone leptonic model, with the X-ray emission evolving from a substantial synchrotron contribution in X1 to increasing IC dominance in X2 and complete IC dominance in X3. The flares are particle dominated, with their energetics driven primarily by changes in the electron population rather than the magnetic field.
\end{enumerate}

These results reveal a systematic evolution of the X-ray flaring in OQ 334, from soft, synchrotron-influenced flares associated with the apparent FSRQ $\rightarrow$ BL Lac transitions to a hard, IC-dominated flare. The X-ray band provides a sensitive probe of the changing balance between the emission components and the CL behaviour.

\begin{acknowledgments}
\textbf{\textit{Acknowledgments}}: This paper is supported by the European Union’s Horizon Europe research and innovation programme under grant agreement No 101131928, project ACME. This work was partially supported by a program of the Polish Ministry of Science under the title “Regional Excellence Initiative”, project No. RID/SP/0050/2024/1.\\
\textbf{\textit{Facilities}}: MAGIC, Fermi (LAT), Swift (XRT and UVOT), ASAS-SN, ZTF, Plank/MMDC. \\
\textbf{\textit{Software}}: Astropy, Gammapy, Fermipy, HEASoft, pyZDCF, agnpy, emcee, iminuit. 
\end{acknowledgments}

\bibliography{references}{}
\bibliographystyle{aasjournal}

\end{document}